# Invisible Users in Digital Health: A Scoping Review of Digital Interventions to Promote Physical Activity Among Culturally and Linguistically Diverse Women

YILIN KE, University of Auckland, New Zealand
YUN SUEN PAI, University of Auckland, New Zealand
BURKHARD C. WÜNSCHE, University of Auckland, New Zealand
ANGUS DONALD CAMPBELL, Hong Kong Polytechnic University, China
MAIRI GUNN, University of Auckland, New Zealand

Digital health has strong potential for promoting physical activity (PA), yet interventions often fail to sustain engagement among culturally and linguistically diverse (CALD) women. Prior reviews focus on short-term efficacy or surface-level localisation, while a design-oriented synthesis of deep cultural adaptation and long-term strategies remain limited. This scoping review systematically screened 1968 records, analysed 18 studies and identified a critical design paradox: techno-solutionist systems overlook social and cultural barriers, while social-support features often fail in low-activity social networks. To address this gap, we propose the Culturally Embedded Interaction Framework, integrating five dimensions: culturally-grounded measurement, multi-modal interaction, contextual and temporal adaptability, embedded social weaving, and theory-guided cultural adaptation. The framework advances beyond accessibility-focused approaches by mapping behavioural theory to design mechanisms that support sustained and culturally plural participation. We provide actionable design principles to help HCI researchers and practitioners move from one-size-fits-all models toward adaptive, theory-informed, and culturally sustaining design.

## 1 Introduction

Digital health technologies (DHTs) are increasingly used to promote health equity and address persistent public health challenges, particularly in PA promotion. Mobile and immersive platforms promise personalised, boundary-less support that can transcend traditional barriers of time, place, and access [17, 37, 40, 50, 52, 54, 71]. For women's health, such technologies have shown potential to increase empowerment and support a range of specific health outcomes [36, 46, 63]. However, from an HCI perspective, this promise is compromised by design misalignments rooted in a techno-solutionist approach that treats cultural and gendered needs as secondary.

Authors' Contact Information: Yilin Ke, yilin.ke@auckland.ac.nz, University of Auckland, Auckland, New Zealand; Yun Suen Pai, yun.suen.pai@auckland.ac.nz, University of Auckland, Auckland, New Zealand; Burkhard C. Wünsche, burkhard@cs.auckland.ac.nz, University of Auckland, Auckland, New Zealand; Angus Donald Campbell, angus.campbell@polyu.edu.hk, Hong Kong Polytechnic University, Hong Kong, China; Mairi Gunn, mairi.gunn@auckland.ac.nz, University of Auckland, Auckland, New Zealand.

**Unpublished working draft. Not for distribution.**

Poorly designed interventions not only limit women's ability to benefit from DHTs but may also exacerbate inequities by embedding cultural and gender blind spots into technical systems [7, 26, 33, 99].

These challenges are particularly acute for women from CALD backgrounds [82], who experience disproportionately high rates of inactivity-related chronic disease [12, 18, 38, 78]. CALD populations are typically defined as individuals from non-English-speaking backgrounds who identify with non-dominant cultural groups in host countries, excluding Indigenous populations who are not recent migrants [30, 65, 75, 82]. For this population, digital health often fails to deliver on its promise. The technology itself may function, but it is built upon design principles that exclude these users by default.

A central reason lies in the dominance of the surface-localisation paradigm [11, 47], which reduces cultural adaptation to technical translation, cosmetic interface changes, or language substitution. This paradigm exemplifies a broader techno-solutionist mindset prevalent in digital health design [5, 67], where complex socio-cultural challenges, such as family responsibilities, gendered norms, community ties, and culturally specific motivations, are reframed as problems solvable through algorithms or interface polish [75, 82].

The limitations of this paradigm become evident when considering how users actually engage with digital systems over time. An influential longitudinal study by Johansson et al. [45] demonstrated that sustained attendance at cultural events is a significant predictor of long-term self-reported health, suggesting that cultural stimulation functions as a "perishable commodity" requiring continual replenishment. Designs rooted in the surface-localisation paradigm privilege short-term usability at the expense of sustained cultural engagement. This mismatch not only drives low participation but also reinforces the very health disparities, such as elevated risks of diabetes and cardiovascular disease [12, 18, 38, 50], that these technologies intend to address. Such neglect has left CALD women structurally invisible within digital health.

Prior reviews in HCI and digital health have examined technology interventions for structurally marginalised populations. Stowell et al. [96] conducted a systematic review of mHealth interventions targeting vulnerable groups, including individuals of low socioeconomic status and racial/ethnic minorities, and found limited gender, cultural, and contextual tailoring, relying heavily on short-term usability metrics rather than examining how experiences evolve over time. Extending this critique, Kouaho and Epstein [48] further argue that wearable and personal informatics interventions overwhelmingly centre on dominant, higher socioeconomic status populations, embedding design assumptions about financial stability, device access, and normative health behaviours that marginalise culturally diverse or resource-constrained groups. Complementing this, Saksono and Parker [90] introduce a socio-cognitive lens emphasising community relationships and social factors in supporting health behaviour, revealing that existing tools often prioritise individualised behaviour change over socially enabled support. However, these HCI reviews remain largely high-level: they provide critical perspectives and call for greater equity but do not engage deeply with the context of a specific vulnerable group, nor do they translate their critiques into operational design guidance.

By contrast, public health research has examined CALD women more directly, documenting the socio-cultural barriers they face in PA participation [18, 19, 93], systematically reviewed intervention effectiveness [29, 75] and highlighted the promise of community-based interventions [14, 65, 103]. However, these reviews tend to be more outcome- and efficacy-oriented, and provide minimal insight into the design mechanisms that facilitate or hinder engagement.

Although these prior reviews provide critical equity-centred analytical lenses and recognise cultural tailoring as essential, a substantial gap remains in how cultural factors are operationalised in design. Existing scholarship rarely examines how cultural or gendered factors can be translated into actionable design paradigms, feature-level decisions, or theory-informed interaction patterns. This absence of implementation-focused guidance leaves a significant disconnect between recognising inequity and designing systems capable of addressing it.

To bridge this gap, we shift the lens from whether interventions work to how their underlying design mechanisms support or undermine engagement. This scoping review systematically examines how digital PA interventions are designed or fail to design for CALD women. Following the PRISMA-ScR methodology [97], we systematically reviewed 18 studies and applied a critical interpretive lens to uncover the dominant design paradigms and their conceptual shortcomings, which limit the cultural and gender responsiveness of digital PA interventions. Our investigation is guided by three research questions:

- RQ1: What design paradigms currently dominate the development of digital PA interventions for CALD women?
- RQ2: What specific design shortcomings undermine their potential for sustained engagement?
- RQ3: How can HCI principles be applied to generate more culturally sustaining designs?

Our analysis reveals a core design paradox: techno-solutionist features routinely ignore relational contexts, while pro-social features prove structurally fragile in low-activity networks. To move beyond this paradox, we introduce the Culturally Embedded Interaction Framework, a five-dimensional model spanning culturally grounded measurement, multimodal interaction, embedded social weaving, theory-informed adaptation, and contextual and temporal adaptability. Importantly, this framework was inductively derived from a systematic analysis of the 18 intervention studies, translating empirical patterns into a structured set of dimensions for culturally embedded interaction. For the HCI community, this paper makes two contributions. First, we offer a critical synthesis of design assumptions in digital health interventions, showing how they preconfigure cultural and gender exclusion. Second, we contribute the Culturally Embedded Interaction Framework, which translates behavioural theory into culturally sustaining interaction patterns for CALD women. These contributions move the field beyond accessibility-centric translation toward design approaches that can better support future genuine co-design efforts [79] and foregrounds cultural pluralism and sustained engagement.

## 2 METHOD

To address our research questions on the design paradigms, socio-technical shortcomings, and culturally sustaining strategies of digital PA interventions for CALD women, we conducted a scoping review [2, 55, 81, 97]. This methodology aligns with our objective to critically examine not merely if interventions work, but how they are conceived, built, and ultimately succeed or fail in their cultural contexts.

Similar to a systematic review, a scoping review examines how research is conducted, summarises the primary focus and findings, identifies conceptual and empirical gaps, and highlights opportunities for future research [15]. Unlike a systematic review, however, a scoping review does not require pre-registration or formal quality appraisal of individual studies [68]. Instead, it emphasises inclusivity across study designs and level of methodological quality, ensuring that early-stage feasibility trials, qualitative studies, and pilot projects are captured alongside randomised controlled trials. This approach is essential for our purposes, as it enables us to synthesise and map the range of design practices in digital health rather than prematurely narrowing the evidence base to a small set of clinically validated interventions.

To ensure transparency and rigour, we adhered to the PRISMA Extension for Scoping Reviews (PRISMA-ScR) [97]. In the remaining subsections, we discuss each of these steps in further detail: (1) literature search strategy; (2) study screening, selection, and eligibility criteria; (3) data charting and analysis; and (4) overview of the Studies.

### 2.1 Literature search strategy

The literature search was conducted in two stages: an initial search (10 May 2025) and an update (14 May 2025). We conducted a systematic search of five major electronic databases: PubMed and Scopus were chosen for biomedical and public health literature, ACM Digital Library and IEEE Xplore for computer science and human-computer interaction research, and Web of Science for multidisciplinary coverage. To address potential delays in database indexing and ensure that we captured relevant digital health research, we supplemented the database search with targeted searches of JMIR Publications, a leading publisher in the digital health domain. To further maximise coverage, we also conducted manual search of citation snowballing and targeted author searches on Google Scholar. These methods continued until no new eligible studies were identified, following established systematic review methodology [43].

Our search strategy was developed through an iterative process of testing and calibration. We initially performed multiple preliminary searches using broad combinations of keywords related to four core conceptual domains: (1) CALD populations (e.g., "immigrant", "refugee", "ethnic minority"), (2) women (e.g., "women", "female"), (3) digital interventions (e.g., "mHealth", "app", "wearable", "virtual reality"), and (4) physical activity (e.g., "physical activity", "exercise", "exergame").

Through several iterations, we refined the strategy to optimise sensitivity and specificity, finalising the search string for each database. The search was temporally bounded to the period from 2015 to 2025, a decade that marks the mainstream adoption of mobile, wearable, and web-based health technologies, to capture the evolution of modern digital health interventions. The complete, reproducible search syntax for all databases is provided in Appendix A.

## 2.2 Paper Screening, Selection, and Eligibility Criteria

The study selection process followed three phases: (1) title and abstract screening, (2) full-text review, and (3) resolution of conflicts through consensus. The first author conducted the initial title and abstract screening of all identified records using Rayyan [1] software, an online systematic review tool designed to manage large-scale screening processes [76, 98]. Rayyan supported the organisation of records, but all final inclusion and exclusion decisions were made manually by the authors. To ensure rigour, the eligibility criteria were iteratively refined in consultation with the co-authors before progressing to full-text review.

The criteria were designed to capture emerging digital health interventions (2015-2025) that addressed PA promotion for CALD women in English-speaking contexts, with an emphasis on culturally informed design rather than clinical rehabilitation. They were organised across five dimensions: population, digital type, target behaviour, study design, and timeframe.

Inclusion Criteria (IC):

- IC1 (Population): Adult CALD women residing in Anglophone countries, where CALD is defined as immigrants, refugees, first-generation (or second-generation with a non-English household language) individuals.
- IC2 (Digital Type): Digital interventions (e.g., apps, wearables, virtual reality/augmented reality) designed to promote physical activity.
- IC3 (Target Behaviour): Studies reporting on physical activity-related outcomes.
- IC4 (Study Design): Empirical research employing quantitative, qualitative, or mixed-methods designs, published in English in peer-reviewed venues.
- IC5 (Timeframe): Studies published between 2015 and 2025.

Exclusion Criteria (EC):

- EC1 (Population): Studies that did not include women; did not include individuals from CALD, immigrant, or ethnic minority backgrounds; or focused exclusively on clinical populations (e.g., cancer rehabilitation, diabetes management), pregnant women, men-only cohorts, children, or adolescents.

---

[1] Rayyan official website https://www.rayyan.ai

- EC2 (Intervention Type): Interventions that did not incorporate any digital or technology-based components, including those delivered entirely face-to-face or through community-based programs without a digital aspect.
- EC3 (Intervention Scope): Digital tools or technologies that were not designed to promote physical activity or exercise, or that were used solely for data collection or monitoring purposes without an integrated behavioural intervention component.
- EC4 (Intervention Focus): Studies where the primary target was disease management or clinical rehabilitation (e.g., cancer rehabilitation, diabetes control, cardiovascular disease recovery) rather than the general promotion of physical activity in healthy or preventive contexts.
- EC5 (Study Design): Non-empirical publications (e.g., reviews, protocols, commentaries, theoretical papers) or studies not published in English.

This process yielded 30 articles for full-text review, which were independently evaluated by two authors. Decisions were recorded in a shared spreadsheet to systematically capture disagreements, which were then resolved through discussion until consensus was reached (Appendix B). Ultimately, 18 studies met all inclusion criteria and comprised the final synthesis corpus. To minimise the risk of missing eligible studies, we also applied manual snowballing and author website checks, which identified four additional publications [10, 57, 60, 64], associated with a multi-phase intervention project (Seamos Activas II). We included mixed-gender CALD samples when women were part of the cohort and the intervention targeted PA. In such cases, we coded design features and engagement outcomes, but did not compute women-only effects unless stratified results were reported. The complete process is illustrated in the PRISMA flow diagram (Figure 1).

### 2.3 Data Charting and Analysis

Data charting was performed to systematically extract and organise key information from the 18 included studies, creating a structured foundation for subsequent analysis. For each study, we documented: (1) bibliographic details (authors, publication year, country of study); (2) methodological attributes (study design, target population characteristics, sample size); and (3) intervention characteristics (digital platform or technology type, specific behavioural strategies employed, cultural adaptation approaches, theoretical frameworks guiding design, physical activity measurement methods, and retention or engagement features). We additionally extracted reported implementation barriers and facilitators, as well as quantitative and qualitative outcomes related to participation and engagement.

Our analytical approach employed a multi-stage process specifically designed to address each research question systematically. The process began with conventional thematic analysis techniques [16] involving iterative coding of extracted data to identify both convergent patterns and divergent findings across studies. Following this process, we coded and categorised the extracted data, and these codes were subsequently organised into higher-order thematic categories explicitly aligned with our three research questions:

For RQ1 (design paradigms), we conducted inductive analysis [3] of design rationales and feature sets to identify and characterise predominant design approaches in the literature.

For RQ2 (design shortcomings), we performed convergent synthesis [41] of articles' limitations of efficacy-engagement disparities to identify structural weaknesses.

For RQ3 (culturally sustaining designs), we employed an interpretive-generative approach [106], identifying positive indicators (elements associated with higher acceptability or retention) to derive design principles and assemble our conceptual framework.

We conducted the mapping of feature frequencies and distributions (e.g., delivery modes, cultural adaptation techniques) to provide additional context for our findings. This mixed-methods approach enabled a comprehensive examination of both the strengths and limitations of current digital approaches, particularly in terms of their capacity to sustain long-term engagement among CALD women.

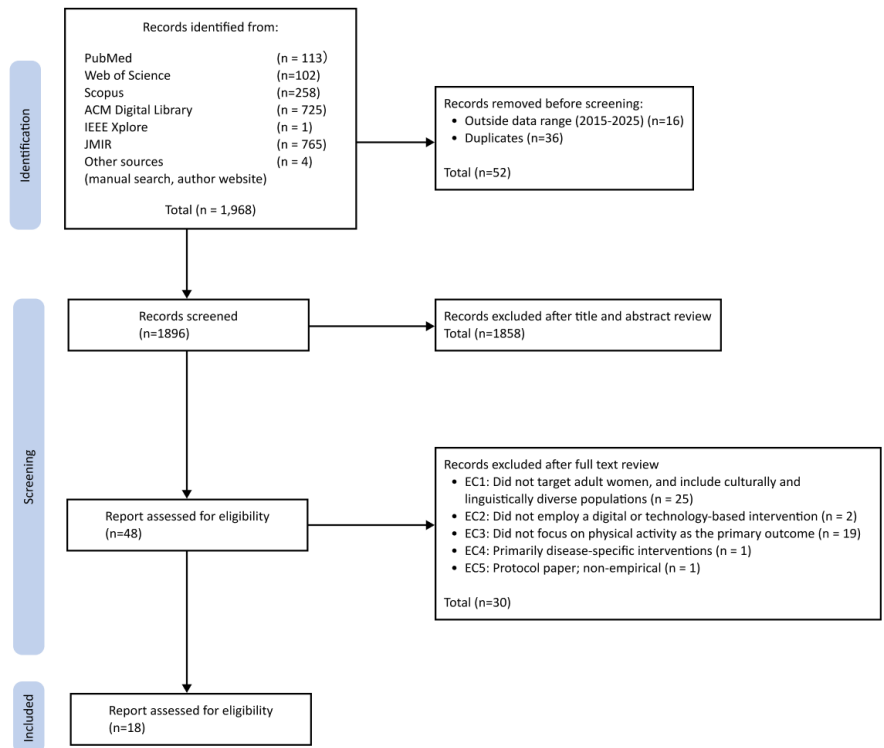

Fig. 1. PRISMA flow diagram of the literature search in this study

### 2.4 Overview of the Studies

The 18 studies reviewed primarily targeted women from culturally and linguistically diverse backgrounds residing in the United States. A majority focused on Latina women (n=14), while a smaller subset centred on Korean American women (n=3) or on midlife Chinese American women (n=1). Only one study engaged a broader sample of multiethnic Asian subgroups, including Chinese-, Tagalog-, and Vietnamese-speaking women (Table 1). The summary of included studies is shown in Appendix C.

Methodologically, the studies spanned diverse research traditions, ranging from randomised controlled trials (RCTs) to formative qualitative inquiry, capturing both exploratory and confirmatory stages of digital intervention development (Table 2). RCTs (n=11) were the most prevalent, including large-scale interventions such as Seamos Activas II and Pasos Hacia La Salud. Pilot and feasibility studies (n=4) evaluated the short-term usability and acceptability of app-based or technology-enhanced interventions. While these offered rich insights into design challenges, their small sample sizes limited broader applicability. Mixed-methods approaches (n=2) integrated quantitative outcomes with qualitative user feedback, enriching evaluation but remaining resource-intensive. In addition, a formative qualitative study (n=1) adopted a user-centred lens to explore emotional, cultural, and contextual factors shaping PA experiences, highlighting the value of early-stage participatory design in culturally sensitive health interventions.

In terms of intended health outcomes, 17 of the 18 studies framed PA as a preventive measure against chronic conditions disproportionately affecting CALD women, such as type 2 diabetes, hypertension, obesity, and cardiovascular disease [61, 86]. This biomedical rationale reflects a dominant orientation across the literature. Only one study centred on psychological well-being, aiming to reduce stress and enhance mood through movement-based engagement.

Table 1. Target population and number of studies

| Target Population | Number of Studies |
| --- | --- |
| Latina Women | 14 [8–10, 23, 33, 51, 57–60, 64, 89] |
| Korean American Women | 3 [21, 22, 92] |
| Chinese American Women | 1 [21] |
| Multi-Asian Women | 1 [71] |

Table 2. Methodology and number of studies

| Methodology | Number of Studies |
| --- | --- |
| Randomized Controlled Trials (RCTs) | 11 [10, 22–24, 51, 57–60, 100] |
| Pilot / Feasibility Studies | 4 [9, 71, 89, 92] |
| Mixed-Methods Approaches | 3 [21, 64] |
| Formative Qualitative Studies | 1 [8] |

## 3 RESULTS

The scoping review synthesises evidence from 18 studies to critically examine the design paradigms underpinning digital PA interventions for CALD women. Our analysis moves beyond efficacy reporting to uncover the assumptions and recurring shortcomings in how these technologies are conceived and built (Appendix D). The findings are structured to directly address our research questions, first revealing the dominant design paradigms (RQ1), then critiquing their inherent limitations (RQ2), and finally identifying seeds for more culturally sustaining approaches (RQ3) (Figure 2).

### 3.1 Paradigms of Cultural Tailoring: From Surface-Level Accommodation to Socio-Cultural Embedding

To address RQ1, we examined how the 18 studies incorporated cultural considerations into the design, content, and delivery of digital PA interventions. Our analysis identified recurring patterns in the types of cultural tailoring used, which we synthesised into two core paradigms (Figure 3): (1) a dominant surface-level tailoring paradigm focused on reducing barriers to accessing digital interventions [8, 21–23, 34, 64, 71, 89, 92], and (2) an emerging deep socio-cultural embedding paradigm, focused on integrating the intervention into the cultural fabric of users' lives [9, 10, 24, 51, 57–60, 100].

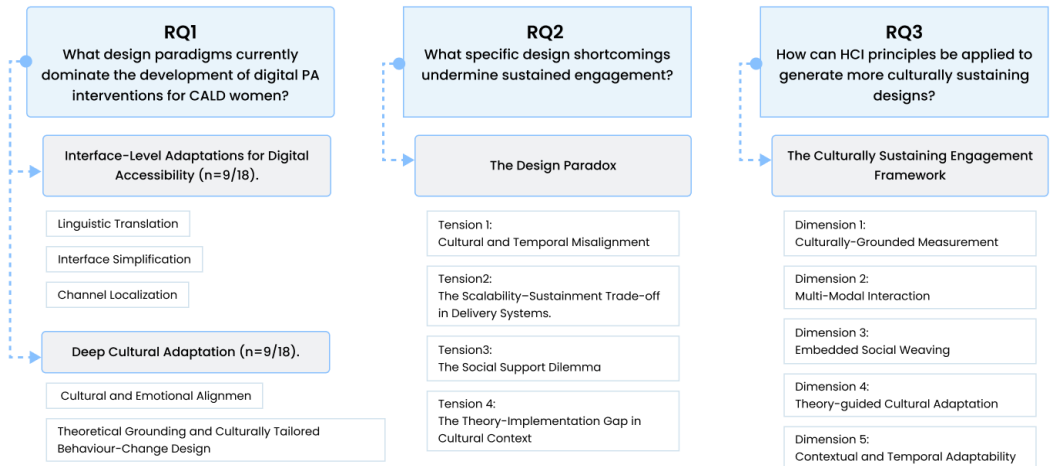

Fig. 2. Overview of the study: from design paradigms (RQ1) to tensions (RQ2) to the Culturally Embedded Interaction Framework (RQ3).

*3.1.1 Interface-Level Adaptations for Digital Accessibility (n=9/18).* Across the review corpus, the prevailing approach to cultural adaptation aligned with Resnicow et al.'s [88] notion of surface-level tailoring, where cultural and linguistic differences were positioned primarily as access barriers rather than as sociocultural structures shaping motivation, identity, or behaviour. Three recurring interface-level strategies defined this paradigm:

(1) Linguistic Translation: Interventions relied on direct translation of text messages, educational content, or website interfaces into participants' primary languages to improve comprehension [8, 21–23, 34, 64, 71, 89, 92].
(2) Interface Simplification: Several studies reduced textual load and readability demands by lowering reading levels, using low-text, visually enhanced, simplified user interfaces to accommodate low health literacy or limited digital skills [8, 34, 71].
(3) Channel Localisation: Interventions often delivered reminders, prompts, and ongoing support through communication channels already embedded in participants' everyday practices, such as SMS, email, or familiar messaging platforms [22, 23, 34, 71, 89, 92].

Across the reviewed studies, such strategies consistently improved accessibility and ease of use. Examples include multilingual systems supported by forward translation and low reading-level messages. Some also incorporated culturally recognisable imagery or voice-to-text features to reduce entry barriers for users with limited digital experience [8, 34, 71]. Korean American interventions similarly integrated culturally familiar communication platforms such as KakaoTalk [2] (widely used mobile messaging in South Korea) and Naver BAND [3] (a group communication app), and recruitment was supported by a community advisory board to strengthen cultural relevance and accessibility [22].

However, the primary outcome of these adaptations was access, not sustained cultural engagement. Success was predominantly evaluated through short-term, technology-legible metrics such as weekly steps, MVPA minutes, and system-use frequency [51, 57, 58]. This pattern aligns with broader critiques in the literature, which suggest that interface-level adaptations can improve initial uptake but are often insufficient for sustaining long-term, culturally situated engagement [70, 94]. The reliance on technology-centric metrics further reflects what Morozov terms a techno-solutionist approach, which reduces complex socio-cultural practices to algorithmically-processable data points [66]. Overall, these adaptations addressed barriers to access rather than embedding cultural values, practices, or motivations into the core behavioural or interactional architecture of the intervention.

---

[2]KakaoTalk official service page: https://www.kakaocorp.com/page/.
[3]BAND official site: https://www.band.us/ko.

*3.1.2 Deep Cultural Adaptation (3 intervention programs; 9 publications).* A minority of interventions are represented by three major multi-phase programs: Pasos Hacia La Salud I [9, 58, 59], Pasos Hacia La Salud II [24, 51, 100], and Seamos Activas II [10, 57, 60], documented across nine publications that have moved beyond the paradigm of surface accessibility. These interventions grounded their behavioural framing, examples, and delivery strategies in Latina women's cultural values, relational identities, and daily lived experiences. Two primary strategies characterised this paradigm:

*(1) Cultural and Emotional Alignment.* Content and communication were designed to resonate with deeply held cultural values and lived realities. Across reviewed studies of the Pasos Hacia La Salud (I/II) and Seamos Activas II programs, cultural adaptation was reflected in the consistent use of Spanish-language delivery and the integration of behavioural examples grounded in Latina women's daily relational and caregiving contexts. Intervention manuals, feedback messages, and activity suggestions frequently framed physical activity through family-oriented routines [24, 58, 59, 100]. Furthermore, several studies document the use of bicultural and bilingual staff and moderated online forums was a documented feature to facilitate culturally and emotionally appropriate communication [24, 59, 100]. These choices supported cultural resonance in both content delivery and interpersonal communication, creating an emotionally supportive context aligned with participants' lived experiences.

*(2) Theoretical Grounding and Culturally Tailored Behaviour-Change Content.* Across the three Latina-focused intervention programmes, behaviour-change theories were primarily adapted through cultural framing rather than structural reconfiguration [9, 10, 24, 51, 57–60, 100]. The primary method of adaptation involved aligning Social Cognitive Theory (SCT) [4] or the Transtheoretical Model (TTM) [83] constructs with culturally salient values and daily circumstances through the use of examples, narratives, and counselling strategies.

A prominent example is the framing of social support through family cohesion, the cultural norm that emphasises collective responsibility and emotional interdependence. In Pasos Hacia La Salud [9, 24, 51, 58, 59, 100] and Seamos Activas II [10, 57, 60], strategies to increase self-efficacy were intentionally embedded in family-centred activity planning, such as dancing with children, walking with extended relatives, or setting shared weekly goals with partners. These strategies aligned SCT constructs with culturally familiar family routines.

A second pattern of deep adaptation involved tailoring SCT constructs related to perceived barriers and environmental constraints. Interventions provided concrete, context-sensitive behavioural solutions, such as walking in groups during daylight hours, selecting indoor or community-centre routes, or integrating PA into childcare and household routines [9, 58].

Finally, the corpus shows that TTM's stage-based tailoring was also applied in culturally specific ways [83]. Rather than offering uniform stage-matched recommendations, manuals and feedback algorithms in Pasos I/II [9, 24, 51, 58, 59, 100] incorporated participants' daily schedules and caregiving routines.

Within these programmes, PA was reframed not merely as an individual health behaviour but as an activity that could enhance caregiving capacity, reduce family burden, and strengthen emotional resilience, aligning motivational pathways with Latina women's relational identities and cultural obligations.

Although nine publications were classified as examples of deep cultural adaptation, they largely report on three closely related intervention programmes: Seamos Activas II [10, 57, 60] and Pasos Hacia La Salud I/II [9, 24, 51, 58, 59, 100]. At the programme level, these interventions demonstrate clear deep cultural adaptation: they are grounded in formative work with Latina women, delivered in Spanish, and explicitly address structural and cultural barriers [9, 24, 57–60]. However, when examined at the level of digital design and interaction features, only a small subset of features, such as family-shared website log-ins or providing separate pedometers for family members, embed cultural values directly into the interaction structure itself [9].

In response to RQ1, our review reveals that digital PA interventions for CALD women are predominantly characterised by a surface-level cultural adaptation paradigm. Most interventions translate content into the target language, simplify wording, and deliver it through familiar channels rather than redesigning the underlying interaction logic or social roles. Although nine publications in our sample qualify as examples of deep cultural adaptation, they mainly represent three closely related Latina programmes (Pasos Hacia La Salud I/II and Seamos Activas II) that share a common, culturally grounded behaviour-change design at the programmes level [9, 10, 24, 57–60, 100], with cultural specificity embedded more in the content and examples than in radically reimagined interaction features.

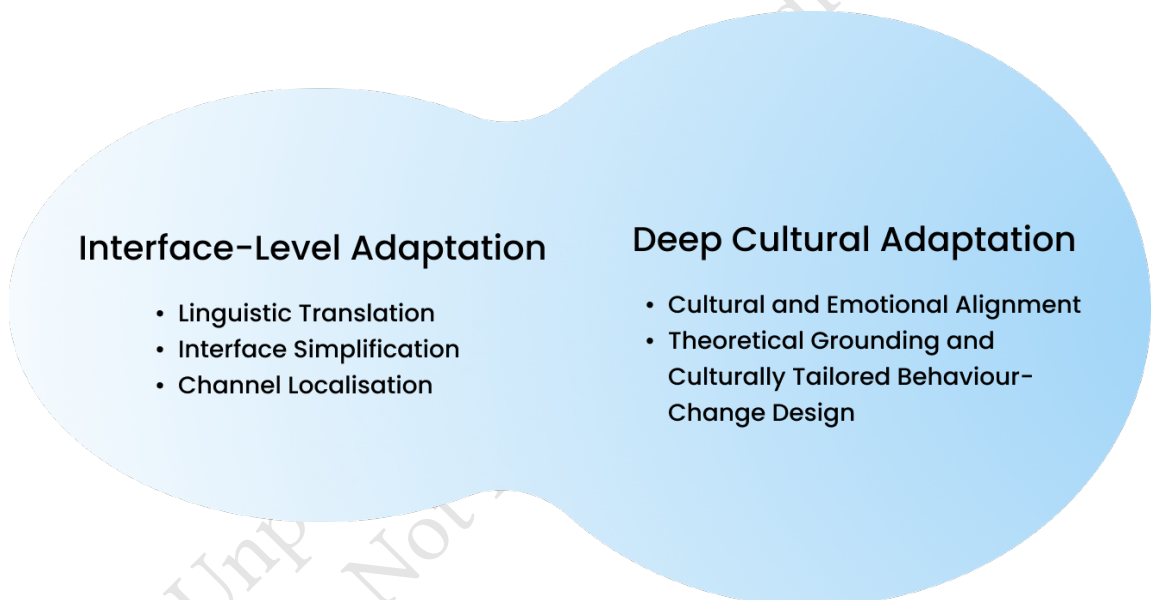

Fig. 3. Cultural tailoring strategies: from surface-level accommodation to deep socio-cultural embedding.

### 3.2 The Design Paradox: Layered Tensions Constraining Engagement

The dominance of the superficially tailored paradigm revealed in Section 3.1 not only leads to ineffective design outcomes but also prevents sustained engagement from the beginning. Our analysis of RQ2 indicates that this structural imbalance generates four interconnected design tensions (Figure 4) that collectively constrain long-term engagement.

*3.2.1 Tension 1: Cultural and Temporal Misalignment (n=4).* Across four studies in our corpus, a recurring tension emerged between how digital interventions were conceptualised and delivered and how CALD women understood health, time, and everyday life [21, 22, 34, 89]. One form of misalignment concerned terminology and conceptual framing. In a pilot study with Chinese and Korean American midlife women, cultural conflicts emerged around the use of terms such as "midlife" and "depression," which carried stigma [21]. In addition, participants' interpretations of what counted as physical activity often diverged from the examples and categories embedded in intervention materials, while inconsistencies in linguistic translation further undermined comprehension and engagement [21].

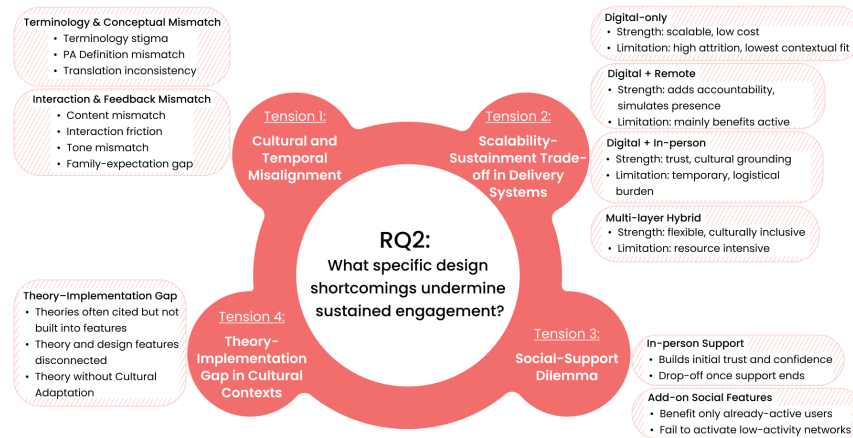

Fig. 4. Four systemic design tensions identified in response to RQ2, illustrating cultural-temporal misalignment, scalability-sustainment trade-offs, fragile social support, and the theory-implementation gap in cultural contexts.

A second form of misalignment concerned friction in everyday interaction with digital tools. In a rural Hispanic context, participants noted that MyFitnessPal did not include common Latino foods, required burdensome portion entries, and faced syncing failures, which collectively reduced motivation and usability [89]. Similarly, in a user-design study of bilingual conversational coaches, the impersonal and culturally irrelevant tone of system feedback posed an additional challenge. Participants who are monolingual Spanish speakers identified mismatches in tone and phrasing, impersonal responses, and the absence of family-inclusive features, indicating that the feedback style did not fully align with their communication expectations or cultural norms around support [34].

Across these studies, recurring cultural misalignments manifested as confusion, frustration, and reduced willingness to continue using the tools, and were described by participants as key barriers to acceptability and sustained use [21, 34, 89]. The resulting friction is not an accidental defect but an inevitable product of a paradigm that prioritises technological efficiency over interaction design rooted in cultural foundations.

*3.2.2 Tension 2: The Scalability–Sustainment Trade-off in Delivery Systems.* The reviewed interventions reveal a structural tension in the design of socio-technical health systems: digital configurations that scale well often fail to sustain engagement, while models that provide sustained, culturally grounded support are difficult to scale. We synthesise the 18 studies into four delivery archetypes that illustrate this scalability-sustainment trade-off.

*(1) Model 1: Digital-only Programmes (n=8).* These interventions delivered PA support entirely via web or mobile channels. Two patterns were observed: (i) web-based, language-localised portals offering tailored content, self-monitoring, feedback, and sometimes discussion forums [8, 9, 51, 58, 59]; and (ii) mobile health apps, often linked with wearable devices such as Fitbit and automated SMS reminders to support adherence [22, 34, 71]. While these models enabled broad reach and low-cost delivery, they consistently faced challenges due to high attrition and low long-term engagement. Across these studies, engagement tended to decline over time, with particularly low participation among women who entered the interventions as completely inactive [58, 59], who remained the most difficult to reach even when interactive features were enhanced.

*(2) Model 2: Digital + Remote Support (n=6).* Digital components were paired with remote asynchronous human support, such as automated SMS for tips or weekly reporting, telephone coaching or check-ins, email feedback, online forums [21–24, 60, 100]. Across these interventions, remote contact was used to simulate social presence, provide accountability, and troubleshoot barriers. However, their effectiveness is highly contingent on participants' pre-existing social capital or network structures and digital communication habits [22, 60]. The engagement data revealed a participation divide. Remote-support features primarily benefited women who were already more active or socially connected, whereas those with weaker or more isolated networks gained limited benefit [22].

*(3) Model 3: Digital + In-Person Support (n=1).* Only one study combined digital tools with structured in-person sessions. In Rowland et al. [89], participants used Fitbit and MyFitnessPal for digital self-monitoring while also attending face-to-face sessions that provided technical assistance, bilingual coaching, and motivational guidance. These in-person components played a central role in onboarding and early engagement, helping participants interpret digital feedback and troubleshoot device–app issues [89]. Adherence data showed that self-monitoring declined when the in-person contact was reduced, and the study reported logistical barriers, including travel time, scheduling difficulties, and staff demand, that constrained scalability [89].

*(4) Model 4: Hybrid Multi-Layer (n=4).* Four interventions combined digital tools with both virtual and in-person support. For example, Shin et al. [92] implemented a programme that integrated multiple layers of human interaction, requiring bi-weekly face-to-face education sessions and monthly small-group walking groups alongside the delivery of weekly reminder text messages and Fitbit tracking to reinforce daily engagement. While these touchpoints offered structure and accountability, some participants experienced pressure and discomfort when discussing adherence [92]. Mendoza-Vasconez et al. combined commercial smartphone apps with an in-person training session to personalise app use, followed by a brief telephone check-in to address technical or behavioural barriers during the maintenance phase [64]. The Seamos Activas II trial further exemplified this hybrid configuration: computer-tailored print materials and automated text messages were integrated with health-coach goal-setting visits and scheduled coaching calls, as described in the design and outcome papers [10, 57]. Across these studies, hybrid delivery offered flexibility to accommodate varying levels of digital literacy and provided culturally grounded support through human facilitation [10, 57, 64]. At the same time, the combination of digital and interpersonal touchpoints highlighted cultural nuances in accountability, showing that while some participants benefited from layered support, others experienced increased social pressure in face-to-face contexts [92].

*3.2.3 Tension 3: The Social Support Dilemma.* Social and peer support are widely recognised as critical in sustaining PA across the included studies [22, 23, 60], especially for CALD women who face intersecting barriers of social isolation, linguistic exclusion, and limited cultural alignment in mainstream health programs [21, 22, 89]. Interventions provided support either virtually, through online peer groups, conversational agents, or web-based coaching [21–23, 34] or through hybrid models that paired digital tools with face-to-face encounters [89, 92]. Across the corpus a core design paradox emerged: digitally added social features rarely activate low-activity networks [21, 22], while in-person support offers only temporary gains and fails to sustain engagement once removed [23].

*(1) Fragility of Add-On Social Features (n=4).* Across four studies, social features embedded as add-ons, including online discussion boards, email coaching, peer groups, and conversational agents, showed limited or uneven engagement. These components were most frequently used by participants who were already active or socially connected, while women with lower baseline activity or weaker social networks engaged less with them.

In Pasos Hacia La Salud, an online message board was provided to enhance peer support; however, usage data revealed minimal visitation, and the feature did not contribute meaningfully to overall engagement with the programmes [51, 58, 59]. Similarly, in a web-based intervention for Chinese and Korean American midlife women, online discussion and email-based coaching attracted little participation despite being framed as culturally supportive mechanisms [21]. Choi et al.[22] further observed that peer-based online groups meaningfully benefited only the most active participants, while women with weaker offline networks, often the intended priority population, derived little benefit.

Conversational-agent-based support exhibited similar limitations. Spanish-speaking women described chatbots as repetitive, technically unstable, and low in social presence, raising concerns about novelty loss and diminishing interpersonal relevance over time [34].

Taken together, these findings indicate that digitally added social features rarely succeed in activating low-activity or socially isolated networks; instead, they often reinforce existing engagement inequalities by primarily serving women who are already active or digitally confident.

*(2) The limitation of In-Person Support (n=3).* Across three interventions that incorporated face-to-face elements, in-person support functioned effectively as an initial scaffold but did not translate into sustained engagement once the contact decreased [89, 92]. In Rowland et al.'s research, bilingual staff assisted with onboarding and troubleshooting, producing short-term increases in engagement. Yet adherence declined steadily over 12 weeks, rising only during face-to-face visits, [89].

A second limitation concerned the way social components intersected with cultural norms and structural responsibilities. In Shin et al. [92], family and work obligations, shaped by traditional gendered expectations, often took precedence over participation in walking groups and scheduled educational sessions. Family members did not consistently encourage physical activity, and the immediate social environment often function as a competing demand rather than a supportive mechanism.

A similar pattern was evident in Chee et al. [21], where weekly one-on-one email coaching was intended to provide accountability, but engagement remained low. Participants were described as "already overwhelmed" by the programme's demands, and the authors noted that social desirability pressures during progress checks may have discouraged responses. In this context, coaching contacts were perceived as burdensome rather than motivating.

Thus, these findings show that the addition of in-person or direct social touchpoints does not automatically enhance engagement. When implemented as frequent progress checks or intermittent support visits, such mechanisms may introduce logistical pressure, cultural tension, or emotional workload that undermines their intended function. Sustained engagement therefore appears to require support structures that are continuous, culturally aligned, and compatible with participants' everyday responsibilities.

*3.2.4 Tension 4: The Theory-Implementation Gap in Cultural Context.* Across the 18 studies, behaviour change theories, primarily SCT and TTM, were widely referenced and consistently used to guide personalised behavioural tailoring. However, their application remained concentrated at the level of individualised content (goal setting, self-monitoring, stage-matched feedback), rather than informing culturally responsive interaction structures or system logic. In practice, theoretical constructs were adapted linguistically or narratively, but rarely reconfigured to reflect cultural norms, family roles, or everyday constraints.

*(1)Theories Driving Behavioural Tailoring.* Ten Latina-focused trials explicitly cited SCT, typically combined with TTM, using these theories to power expert systems, tailoring algorithms, or motivational scripts [9, 10, 24, 51, 57–60, 64, 100]. Five additional studies implemented SCT-related components, such as self-monitoring or feedback, without naming the theory explicitly [21–23, 89, 92]. These constructs were operationalised through standard behaviour-change mechanics. Pasos Hacia La Salud I used an expert system to generate tailored reports on activity levels, goals, and perceived barriers [58]. Enhanced versions (Pasos II; Seamos Activas II) added web-based logs, progress graphs, and interactive SMS reporting to support ongoing self-monitoring [10, 57]. Commercial tools such as Fitbit and MyFitnessPal automated self-monitoring in studies with Korean American and rural Hispanic adults [89, 92].

Additionally, TTM was referenced in 11 studies [9, 10, 22, 24, 51, 57–60, 64, 100]. In the Latina programmes, TTM stages of change were used to structure intervention tailoring: baseline and follow-up stage assessments informed the frequency, intensity, and type of materials delivered [9, 10, 57–60, 64]. In contrast, in the Korean American pilot, TTM was used to describe participant readiness for activity and to determine eligibility but was not integrated into digital tailoring [22].

*(2) Behaviour Change Theories Used Without Systematic Cultural Adaptation.* Despite extensive behavioural tailoring, SCT and TTM were seldom embedded into culturally specific interaction structures. For example, SCT's social support construct was often implemented through generic prompts to seek encouragement from others, or via anonymous message boards, rather than through mechanisms explicitly aligned with family norms. In Pasos Hacia La Salud, the use of the online message board was low and not associated with higher MVPA [9]. Similarly, in technology-enhanced programmes for Korean American adults, multi-source support strategies did not produce significant changes in perceived support from family or friends [92].

Cultural interpretations of PA further constrained theoretical alignment. In formative work with Chinese and Korean American midlife women, participants viewed many daily activities as already "sufficient," creating misalignment with MVPA-oriented goal-setting frameworks; stigma linked to labels such as "midlife" reduced willingness to enrol or remain engaged [21]. App-based interventions for multiethnic Asian adults emphasised language matching and culturally familiar examples for usability, yet retained generic self-monitoring and goal-setting mechanisms [71].

Only a minority of studies have demonstrated the effective translation of theory through socio-technical systems that embed theoretical constructs within culturally responsive interaction loops. The Seamos Activas II series [10, 57, 60] systematically mapped SCT and TTM constructs into tailored text messaging, motivational interviewing, and contextual feedback, demonstrating that even automated low-burden, theory-driven interaction features, such as individualised goal setting and personalised feedback, significantly predicted sustained MVPA at 12-month follow-up [57].

Overall, across the corpus, behaviour-change theories were applied consistently, but primarily to support personalised behavioural content rather than culturally adaptive interaction design.

In direct response to RQ2 (What specific design shortcomings undermine sustained engagement?), our analysis identifies four systemic tensions in the design of digital PA interventions for CALD women: (Tension 1) cultural and temporal misalignment, (Tension 2) the scalability-sustainment trade-off in delivery systems, (Tension 3) the social-support dilemma, and (Tension 4) the theory-implementation gap in cultural context. These shortcomings do not exist in isolation but are interdependent, collectively constraining the capacity of current digital interventions to support sustained, culturally situated engagement.

### 3.3 Toward a Culturally Embedded Interaction Framework

Our analysis of the 18 interventions revealed a dispersed set of HCI-relevant strategies, spanning multimodal delivery, cultural usability, social embedding, theory-informed tailoring, and temporal adaptability, that recurred across the reviewed studies. To address RQ3 (How can HCI principles be applied to generate more culturally sustaining designs?), we integrated these empirical patterns through a three-step synthesis process, culminating in the identification of five core design dimensions.

*Step 1: Extraction HCI-relevant design strategies across interventions.* We first systematically extracted concrete design practices that were explicitly reported to enhance engagement, acceptability, or cultural relevance. This yielded a catalogue of recurrent strategies, including: prioritising cultural usability through low-text, visual-first interfaces [8, 9, 21, 71]; employing multimodal interaction across SMS, web, apps, and wearables to meet diverse user needs [9, 23, 71, 89, 92]; structuring social support through promotoras, bilingual coaches, or familiar platforms [8, 21, 22, 34, 89]; and implementing theory-driven tailoring combined with maintenance scaffolds such as stage-matched messages and adaptive feedback [9, 10, 24, 51, 57, 58, 60, 64, 100].

*Step 2: Mapping strategies to design tensions.* We then analysed how these strategies corresponded to the four design tensions in Section 3.2. For instance, the cultural-temporal misfit (T1) can be addressed through dimensions of culturally grounded measurement and theory-informed cultural adaptation, implemented via orientations such as localising language and imagery, embedding culturally familiar channels, and avoiding stigmatising labels [9, 21, 71, 92]. Strategies responding to the scalability-sustainment trade-off (T2) frequently focused on reducing interaction burden, leading to orientations such as minimising manual logging and lowering technical barriers [23, 51, 57, 71, 89]. Culturally embedded relational structures, such as promotoras or culturally familiar social platforms, responded to the social-support dilemma (T3) by providing trusted, culturally legible pathways for connection [8, 22, 34, 89]. Finally, culturally framed theory implementation and stage-tailored maintenance phases addressed the tension of behaviour-change theories being used without systematic cultural adaptation (T4) [10, 21, 24, 57, 60, 64, 92, 100].

*Step 3: Integrating mapped strategies into a five-dimension design model.* Through an iterative process of clustering the mapped strategies based on thematic similarity and their shared focus on interaction design, five distinct dimensions consistently emerged: Culturally-Grounded Measurement, Multi-Modal Interaction, Embedded Social Weaving, Theory-Guided Cultural Adaptation, and Contextual and Temporal Adaptability. This mapping, detailed in Table 3, illustrates how each dimension integrates multiple strategies and addresses specific tensions.

### 3.4 The Framework Dimensions

The framework (Figure 5) consists of five interconnected dimensions. It is not a static checklist but a dynamic system where culture guides design and is reinforced through ongoing engagement. All examples in this subsection are drawn directly from the 18 interventions in our review corpus.

*Dimension 1: Culturally-Grounded Measurement.* This dimension advocates for measurement practices that extend beyond device-based metrics to incorporate culturally or psychosocially relevant indicators. For example, trials with Korean American women combined wearable-derived walking data with self-efficacy and perceived social support, documenting that increases in activity did not consistently correlate with higher felt encouragement from family or friends, a nuance invisible to step counts alone [22, 92]. App-based programmes for Chinese-, Tagalog-, and Vietnamese-speaking adults integrated Fitbit steps with multilingual education modules and participant feedback on feasibility and cultural fit, foregrounding whether content felt usable and relevant in daily life [71]. Work with rural Hispanic adults similarly triangulated technology logs with usability and reports of cultural mismatch, such as the absence of a Latino food database [89]. Formative and pilot studies with Chinese and Korean midlife women documented how labels like "midlife" and mental-health framing affected acceptability, underscoring the need to measure stigma, role expectations, and family obligations alongside PA behaviour [21]. In addition, conversational-agent work with English- and Spanish-speaking women elicited qualitative indicators such as warmth, empathy, and family inclusivity that function as cultural performance metrics for "humanness" in coaching, not captured by devices [34]. These empirical examples reinforce Lupton et al.'s observation that quantitative data alone are insufficient to represent the complexity of lived experience and the body [53]. Collectively, these studies illustrate a measurement repertoire that couples objective PA with culturally salient constructs, thereby expanding the criteria for evaluating effectiveness.

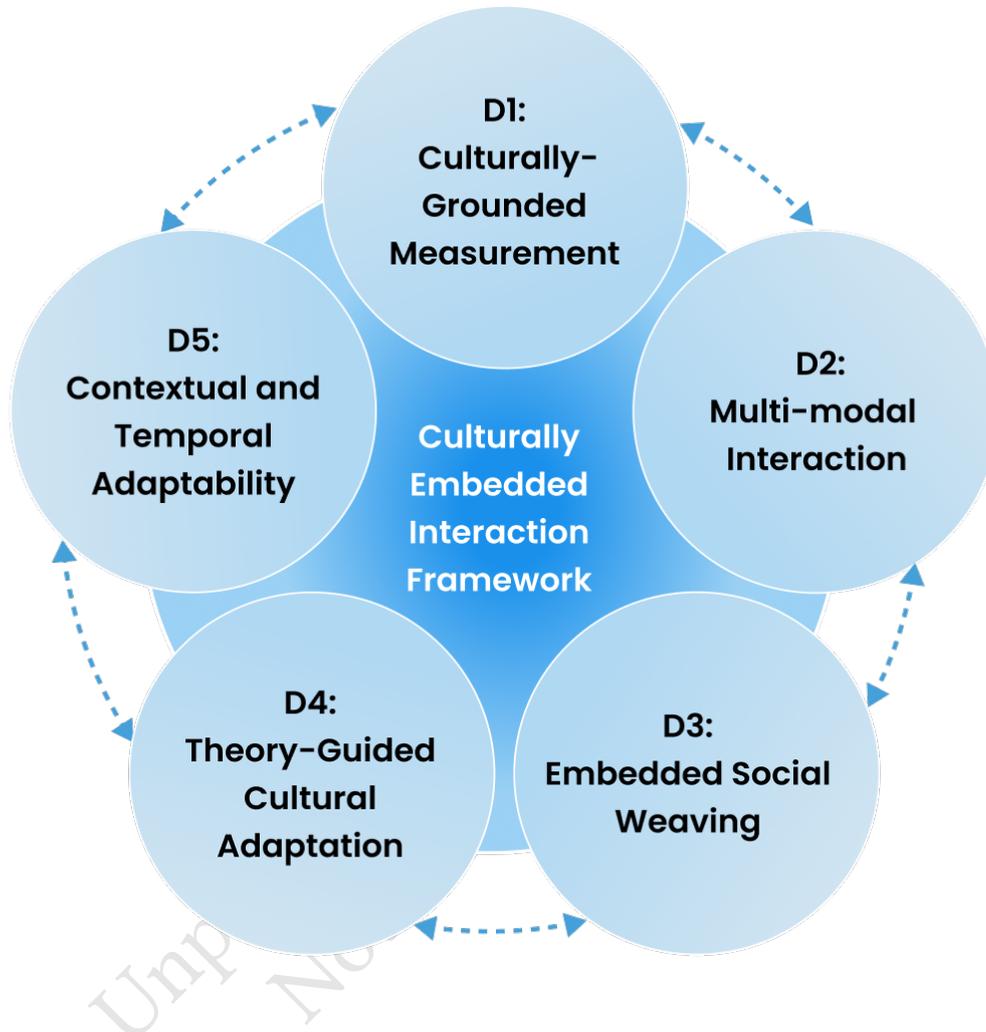

Fig. 5. Proposed Culturally Embedded Interaction Framework: consisting of five interconnected dimensions: culturally-grounded measurement, multi-modal interaction, embedded social weaving, theory-guided cultural adaptation, and contextual and temporal adaptability. The framework illustrates how culture functions as a dynamic driver of digital PA intervention design, continuously shaping and reinforcing sustained engagement.

*Dimension 2: Multi-Modal Interaction.* Evidence indicated a recurrent pattern of employing multiple communication channels to accommodate diverse literacy levels, device access, and communication practices. Several design-phase studies with Latino communities reported that low-text, visual-first interfaces, such as images, icons, and short videos, were easier to understand and more motivating than text-heavy screens [8]. In a bilingual chatbot study, Spanish-speaking women expressed preferences for simplified conversational flows and voice-to-text input, indicating the usefulness of audio-based interaction in reducing literacy demands [34]. Among rural Hispanic adults, participants recommended voice guidance, streamlined flows, and fewer manual entries when app and device complexity hindered continued use [89].

Multi-channel programs that paired apps with short, readable text messages and brief videos were used to improve reach and reinforcement. An intervention for multi-Asian-language speakers coupled culturally adapted app content with daily SMS nudges [71]. Similarly, Spanish-language website interventions integrated SMS reminders to sustain engagement [9, 51, 57]. Randomised trials with Latino adults demonstrated the combined use of text messaging and wearables, where low-reading-level SMS prompts supported day-to-day activity tracking [23, 57].

*Dimension 3: Embedded Social Weaving.* Across the reviewed studies, social connection was implemented through culturally familiar and trusted relational structures rather than generic community features. Several interventions incorporated culturally matched intermediaries, such as promotoras, bilingual coaches, or community advisory boards, which facilitated participant engagement and lowered access barriers [8, 21, 89].

Studies showed differential engagement based on social platform design. Korean American women who actively used a familiar social app showed higher activity than low-engagers, though access alone was insufficient to ensure social support [22]. Conversely, when "social" features are inadequately designed, such as message boards, non-customisable chats, or lacking peer pathways, participants were observed to disengage or request clearer, culturally resonant connection methods [9, 34, 71].

Multiple trials further documented that increases in PA did not consistently correlate with increased perceived support from friends or family, and that rigid, evaluative check-ins sometimes negatively affected engagement [21, 92]. Overall, the evidence suggests that effectively supporting CALD women digitally involves weaving intervention features into the fabric of existing, trusted community structures, rather than deploying standalone social tools.

*Dimension 4: Theory-guided Cultural Adaptation.* Across the reviewed studies, behavioural theories such as the TTM and SCT were frequently referenced, but their translation into concrete design features varied substantially. For instance, in Seamos Activas II, the intervention drew on both the TTM and SCT. TTM informed the stage-tailored print and text-message content, while SCT informed the use of goal-setting, self-monitoring, and feedback, which together helped sustain activity levels up to 12 months [57, 60]. Other interventions, however, invoked theory primarily on a conceptual level with limited corresponding feature-level implementation [10].

Evidence indicated that cultural framing influenced the application of theory-driven techniques. Research with Korean American women found goal-setting and self-monitoring to be more effective when delivered through a familiar social platform [22]. Latina participants reported greater motivation when prompts and feedback referenced family roles or collective values, rather than solely individual progress [34, 89]. Similarly, app-based interventions for multiethnic Asian groups indicated that language matching and culturally specific exercise examples were essential for enabling participants to enact SCT-derived strategies [71]. These studies demonstrate that theory-informed design is most effective when adapted to cultural contexts.

*Dimension 5: Contextual and Temporal Adaptability.* Across several interventions, engagement patterns were influenced by temporal and contextual factors documented in the studies. Cultural calendars affected participation flows: recruitment, program delivery, and user engagement decreased around major cultural holidays such as Lunar New Year and U.S. Thanksgiving [21]. Research also revealed episodic engagement patterns, where participation frequently peaked following scheduled contacts and declined during interim periods [21, 89].

Several interventions employed either rhythm-based or real-time support approaches. In a Korean online community program, weekly challenges and feed updates were associated with consistent platform use [22]. In Seamos Activas II, stage-tailored text messages, brief check-ins, and periodic feedback reports contributed to supporting activity levels over a 6-month period [57, 60]. Correlational analyses linked platform activity duration to behavioural outcomes, with higher and sustained use of intervention websites predicting greater levels of moderate to vigorous PA [51]. Maintenance-focused pilots further documented that low-burden supports, such as automated feedback loops and occasional tailored prompts, were associated with continued activity after program completion [60, 64]. These studies show how temporal patterns, routine alignment, and culturally shaped schedules influenced ongoing engagement.

Our analysis of 18 digital PA interventions reveals a field characterised by two cultural design paradigms and four systemic tensions. The inductive synthesis of strategies across this corpus yielded five interconnected dimensions, establishing an empirically grounded foundation for understanding culturally embedded interaction design in digital health.

Table 3. Mapping design tensions to framework dimensions

| Design Tension | Framework Dimension(s) | Design Orientation |
| --- | --- | --- |
| T1 Cultural-Temporal Misfit | D1: Culturally-Grounded Measurement | Localise language and imagery [8–10, 21, 22, 24, 51, 57–60, 64, 100] |
| | D4: Theory-Informed Cultural Adaptation | Use culturally familiar channels [8–10, 21, 22, 24, 33, 51, 57, 58, 71, 89, 100] |
| | | Avoid stigmatising labels [21] |
| | | Make content culturally concrete [22, 33, 71, 89] |
| | | Keep text simple / visual-first [8–10, 23, 33, 57, 58, 71, 89] |
| T2 Scalability-Sustainment Trade-off | D2: Multi-Modal Interaction | Offer multi-modal access [10, 22, 23, 57, 71, 89] |
| | D5: Contextual and Temporal Adaptability | Lower technical barriers [8, 21–23, 33, 71, 89] |
| | | Minimize manual logging [9, 24, 51, 58, 60, 89] |
| T3 Fragile Social Support | D3 Embedded Social Weaving | Embed culturally matched communities [22, 33, 71, 92] |
| | | Leverage trusted community brokers [8–10, 21, 22, 24, 51, 57, 58, 60, 89, 100] |
| T4 Theory-Implementation Gap | D4: Theory-Informed Cultural Adaptation | Embed SCT/TTM into design features [10, 22–24, 51, 57, 58, 60, 64, 89, 100] |
| | | Sustain with "light-touch" scaffolds [10, 24, 51, 57, 58, 60, 64, 92] |
| | | Co-design cultural transcreation [8, 21, 22, 34, 71, 89] |

## 4 DISCUSSION

To our knowledge, this is the first scoping review to critically examine how digital technologies have been designed to enhance and sustain PA among women from CALD backgrounds. Our synthesis reveals both the promise of existing interventions and their persistent shortcomings. In this section, we revisit our three research questions and articulate how our findings extend knowledge in HCI.

### 4.1 The Paradox of Cultural Design in Digital Health

In addressing RQ1 and RQ2, our findings reveal a structural design paradox that undercuts long-term engagement in digital PA interventions for CALD women. On one hand, techno-solutionist features, automated reminders, algorithmic tailoring, and simplified interfaces, treat culture primarily as an access problem and overlook the relational, cultural, and temporal contexts shaping everyday PA practices [10, 21, 58, 71, 89, 92]. On the other hand, pro-social features intended to foster motivation frequently collapse in low-activity or socially isolated networks, benefiting only those who are already active or digitally confident [22, 34, 51, 92]. These patterns reveal that disengagement is not incidental but systemic, rooted in universalist design logics that privilege efficiency over cultural and social ecologies [6, 44].

*From Surface-Level Localisation to Cultural Reductionism.* Although cultural tailoring is widely acknowledged across interventions, its implementation is overwhelmingly constrained to surface-level localisation through translation, simplified readability, familiar messaging channels. This approach fails to embed deeper cultural values, relational norms, or communication styles into the interaction architecture itself [8, 9, 22, 23, 34, 51, 57–60, 64, 71, 92, 100]. Even when studies incorporated cultural brokers such as promotoras, bilingual coaches, and community advisory boards, who played important roles in ensuring cultural resonance during recruitment or initial content development [8, 21, 22, 71, 89]; their role was typically confined to the initial phases, such as recruitment or content development. Consequently, their deep cultural expertise was not translated into the system's ongoing interaction logic, leaving culturally grounded relational mechanisms absent from its core behavior. This disconnect reflects what Resnicow et al. describe as surface-structure adaptation, reducing cultural practices to aesthetic markers while leaving underlying relational and motivational logics untouched [88].

*Limitations of Techno-Solutionist and Social Features.* Efficiency-driven features such as automated reminders or algorithmic personalisation were widely deployed but did not account for the relational and motivational factors identified in the corpus [57, 71, 89].

Conversely, "add-on" social features, such as online forums or chat-based encouragement, assume that digital connectivity can activate otherwise dormant social networks. Our synthesis demonstrates the opposite: these features disproportionately serve individuals who are already socially resourced or physically active, while those with weaker networks, often the primary target of interventions, receive limited benefit [22, 34, 51]. This ineffective layering of social features exposes a second critical flaw: the failure to account for the culturally variable nature of social support itself. Family involvement likewise produced divergent effects across cultural groups: it functioned as reinforcement in some contexts, such as Latina women who welcomed family-oriented framing [58, 64], but acted as a burden or source of pressure for others, including Korean American women who preferred independent or private modes of engagement [21, 92]. The resulting pattern is clear: neither automated technical features nor generic social add-ons meaningfully map onto the relational infrastructures that mediate CALD women's daily lives.

*Cultural Embedding and Maintenance.* A related pattern concerns the temporal dimension of cultural design. Culture and maintenance were often treated as secondary or optional components, rather than as structuring principles that inform how support should evolve and adapt over time. This explains the recurrent finding of steep declines in engagement after an initial active phase, even when early behavioural gains are achieved [10, 51, 89, 92].

By contrast, the few studies that embedded cultural logic into their temporal structure demonstrated sustained impact. For example, interventions that used stage-tailored prompts and periodic feedback reports within a culturally coherent narrative showed adherence extending to 12 or even 24 months [60, 100]. Lightweight, culturally-framed automated prompts similarly supported engagement beyond the formal intervention period [57, 64]. These divergent outcomes reveal a fundamental insight: long-term engagement failures are not a problem of reminder dosage, but of temporal-cultural misalignment. Successful maintenance requires designing follow-up support so that cultural logics, relational structures, and daily rhythms are woven into the fabric of ongoing interaction, not appended as an afterthought.

*Limitations of Human-Centred and Participatory Approaches in Existing Interventions.* The prevalent pattern of surface-level cultural adaptation can be attributed, in part, to fundamental limitations in how human-centred and participatory principles are translated into practice. A critical observation is that none of the reviewed studies explicitly situated their methodology within established frameworks of Human-Centred Design (HCD) [101], User-Centred Design (UCD) [56], or Participatory Design [84]. Our classification, therefore, reflects the methods that appeared, such as cognitive interviewing, usability testing, advisory consultation, or co-design workshops.

This distinction reveals a structural gap in the current digital health landscape: while many interventions conduct user-informed formative research, few extend these insights into the iterative co-creation of interaction features, social mechanisms, or long-term maintenance structures.

UCD or HCD methods were most visible during early formative research. Studies such as Seamos Activas II and Pasos Hacia La Salud employed focus groups, cognitive interviews, and literacy assessments to establish linguistic clarity, identify cultural barriers, and ensure content comprehensibility [10, 58]. Similarly, studies with Asian and Hispanic populations employed usability interviews to develop low-text interfaces and culturally resonant content [8, 71]. While these activities reflect the core UCD or HCD principle of understanding user context, their impact remained constrained. The design changes they prompted largely focused on improving content readability and surface-level usability, failing to inform the deeper interaction logic or behavioural scaffolding that characterises sustained engagement [51, 57].

Participatory approaches were present but limited in scope, primarily through advisory structures rather than direct co-design. Korean American and rural Hispanic interventions engaged community advisory boards to review recruitment materials, communication tone, and delivery channels [22, 89]. While these forms of stakeholder engagement shaped programme framing and cultural positioning, they did not translate into the systematic co-creation of digital features or behavioural scaffolding. Only one study extended Participatory Design into the design of interaction behaviour by conducting co-design workshops [34].

Overall, while user or human-centred and participatory methods generated important cultural and contextual insights during formative phases, these insights seldom informed the design of sustained, interaction-level mechanisms. This disconnect highlights a key opportunity for HCI: to integrate user-centred and participatory processes throughout the research and design lifecycle, ensuring that cultural understanding not only shapes content but is also structurally embedded in how digital systems behave over time.

### 4.2 Culturally Embedded Interaction Framework: A Design Framework for Sustained Participation

In response to the structural paradox revealed in our review, where techno-solutionist features overlook cultural and relational contexts, while social features collapse in low-activity networks, we propose the Culturally Embedded Interaction Framework. This framework represents more than a checklist; it proposes a fundamental reorientation for digital health design: from achieving short-term, cross-cultural accessibility to fostering long-term, culturally sustainable participation.

Consistent with HCI traditions that position frameworks as interventions into design practice rather than descriptive summaries [6, 31], the framework integrates empirical evidence with culturally situated design theory to guide both analysis and generative design. It serves a dual, critical function: as an analytical lens to diagnose the systemic failures of surface-localization, and as a generative guide for creating new, culturally resonant designs. It is structured around five interdependent dimensions that address the shortcomings of the dominant surface-localisation paradigm, as detailed in Table 4.

The framework moves beyond treating culture as a barrier to be translated, and instead operationalises cultural logics as generative resources for interaction design. For instance, drawing on evidence from long-term trials [60, 100] on the importance of sustained support, the framework argues for designing long-term socio-technical infrastructures [1] that integrated continuous support, adaptation, and integration into everyday practice. This shift is achieved through dimensions of contextual and temporal adaptability (D5), which prioritise continuity and resilience [13, 25, 105] and Embedded Social Weaving (D3), which establishes durable, culturally grounded relational structures to sustain engagement over time.

The framework provides distinct entry points for diverse stakeholders engaged in designing and delivering culturally grounded digital health systems. For researchers, it offers a structured, culturally adjusted lens for identifying challenges, mismatches, and support expectations in early formative work, and for examining how behaviour-change theories operate across different cultural contexts [6, 28, 44, 104]. It can guide sampling strategies and the development of culturally sensitive interview and survey protocols, foregrounding whose voices and value systems are centred in the design process [31, 84]. The framework also supports systematic evaluation of emerging intervention concepts against its five dimensions, helping researchers diagnose why prior interventions struggled and design studies with stronger cultural validity and more interpretable outcomes [13, 28, 94].

Table 4. Framework dimensions and their purposes

| Dimension | Purpose |
| --- | --- |
| D1: Culturally-Grounded Measurement | To ensure tracking and feedback resonate with culturally specific values and definitions of health and activity. |
| D2: Multi-Modal Interaction | To meet users across their preferred channels and literacy levels, ensuring the intervention fits seamlessly into their daily practices. |
| D3: Embedded Social Weaving | To architect bridging social capital by actively designing opportunities for new, meaningful connections to form, rather than relying on pre-existing networks. |
| D4: Theory-Guided Cultural Adaptation | To transcreate behavioural theory by deeply re-engineering motivational constructs to align with cultural worldviews, moving beyond superficial translation. |
| D5: Contextual and Temporal Adaptability | To design for life's rhythm by creating interventions that are fluid and long-term. |

For designers, the framework translates empirical insights into actionable design requirements throughout the lifecycle, from requirements gathering to prototyping and iteration, extending long-standing commitments to situated and value-sensitive design [6, 28, 31]. The framework also provides criteria for evaluating design alternatives, such as whether interaction models reflect collectivist orientations, temporal rhythms, or culturally familiar communication patterns, helping designers move beyond surface localisation toward deeper cultural embedding [74]. For community health workers and cultural brokers, the framework clarifies how their relational expertise can shape the system's social fabric and maintenance patterns, moving their role beyond recruitment toward co-creating culturally grounded socio-technical infrastructures. This aligns with HCI work that highlights the importance of intermediaries in sustaining engagement within complex social systems [27, 44].

### 4.3 Future Validation of the Framework

Although the framework is grounded empirically in 18 interventions, its validity must be established through situated enactment rather than universal generalisation [49]. Following HCI traditions that position frameworks as forms of intermediate-level knowledge [42], the Culturally Embedded Interaction Framework should be evaluated by its usefulness in guiding reflection, shaping design decisions, surfacing cultural tensions, and supporting richer explanations of engagement. Future validation should prioritise contextual, socio-technical inquiry over metric-driven evaluation.

Future work can pursue several complementary pathways that align with Research through Design [106], Research-Informed Design [80], and established framework-refinement practices in HCI [84, 101]. First, participatory and co-design studies can use the framework as a structuring device within Research-Informed Design processes. Building on established HCI traditions of participatory design or community based participatory research with marginalised communities [49, 85], facilitators can incorporate the five dimensions into activities such as card sorting, scenario building, journey mapping, and speculative prototyping. Working with CALD women, family members, and cultural brokers, these sessions can examine whether the dimensions effectively support participants in articulating cultural needs, relational responsibilities, emotional barriers, and temporal rhythms that are often flattened in conventional requirements-gathering or behaviour-change surveys. Within a Research-Informed Design workflow [80], the framework becomes a generative scaffold: it shapes design conversations, focuses the interpretation of qualitative insights, and orients ideation toward culturally embedded interaction mechanisms.

Second, the framework can be applied as a heuristic and analytic lens within Research through Design cycles [106] for evaluating existing or newly developed digital interventions [72]. In this evaluative mode, the five dimensions provide high-level prompts that draw evaluators' attention to cultural mechanisms often overlooked in conventional assessments of digital health systems and can help identify where culturally embedded interaction is present, partial, or absent, and how such gaps may help explain patterns of uptake, appropriation, or attrition. As a pathway for future validation, the framework can be applied when reviewing existing PA technologies or evaluating early design concepts to see whether it reveals cultural blind spots or mismatches that would otherwise be missed.

Third, longitudinal qualitative studies can examine how designs informed by the framework sustain engagement over time [13, 45, 105]. Integrating the framework into a full Research through Design cycle involves using it to guide prototype development, deploying these prototypes in real-world contexts, and iteratively refining both the artefacts and the framework itself as misalignments emerge between expected and observed practices [106]. As a concrete validation pathway, the framework could be applied to guide the design of a small prototype or a revised feature, which is then deployed for several weeks with CALD women. By following participants' evolving interactions, reflections, and adaptations across this period, researchers can assess whether elements derived from the five dimensions influence durability of use, perceived cultural fit, or ongoing motivation.

Given the situated and heterogeneous nature of cultural practices, validation should rely on context-specific deployments rather than attempts at universal generalisation [11, 27, 28, 88].

### 4.4 Recommendations: Design Implications for HCI Research

The Culturally Embedded Interaction Framework provides a theoretical strategy for resolving the design paradox. Here, we translate its five dimensions into four core design implications, actionable guiding principles for researchers and practitioners to create design choices for digital PA interventions targeting CALD women (Appendix E).

*Implication 1: Design for Temporal Fluidity and Long-Term Maintenance.* The framework calls for a shift from designing time-bounded interventions toward creating socio-technical infrastructures that can accommodate the fluctuating temporalities of users' lives. This reframing positions maintenance not as a post-intervention add-on but as an organising principle that shapes how a system evolves with users over months or years.

Digital interventions require embedding ongoing low-burden, long-term scaffolding mechanisms such as stage-tailored messages, adaptive feedback loops, and periodic check-ins that sustain motivation without overwhelming users. This perspective foregrounds the idea that long-term engagement is sustained not through escalating demands but through infrastructures that remain present, adaptable, and culturally resonant over time.

For HCI research, this implies treating temporality itself as a core design material, one that requires attention to the durability, adaptability, and cultural pacing of digital systems, rather than their short-term behavioural effects.

*Implication 2: Embed Behavioural Theory into Interaction Features.* Across the reviewed studies, behavioural theories such as SCT and TTM were frequently cited, but they rarely shaped the actual interaction design. In most interventions, theory served as a conceptual justification rather than a driver of specific features, resulting in limited cultural relevance and weak long-term engagement. Here, interaction features refer to the concrete system behaviours, such as prompts, feedback loops, goal-setting mechanisms, or social reinforcement pathways, through which users experience and enact behaviour-change processes.

For HCI, this signals the need to shift from referencing theory to translating theory into culturally aligned interaction patterns. Constructs such as readiness, self-efficacy, social reinforcement, and habit formation cannot be implemented as universal mechanisms; they must be adapted to the cultural contexts in which behaviour change unfolds. For example, confidence may be individually framed in some populations yet relationally or collectively constructed in others; likewise, the motivational force of goals or feedback depends on how responsibility, obligation, and encouragement are culturally understood.

This reorientation positions theory as a practical design resource rather than a static citation, enabling digital PA systems to support behaviour change that is both theoretically grounded and culturally resonant.

*Implication 3: Rethink the Role of Social Support.* Evidence from the corpus indicates that presuming the existence of supportive, high-activity networks leads to fragile social features that fail in low-energy social contexts [22, 92]. This pattern echoes prior HCI evidence showing that social features, such as sharing or social comparison, do not motivate all users uniformly; while they can support already engaged individuals, others experience pressure or reduced motivation, highlighting the need for personalised rather than generic social mechanisms [62, 69].

By contrast, studies that engaged trusted intermediaries, such as promotoras, bilingual coaches, or community advisory boards, demonstrated stronger cultural resonance and acceptability [10, 71, 89]. In these programmes, social support was not assumed to emerge organically from peer networks but was deliberately mediated through actors who understood participants' language, cultural norms, and everyday constraints.

For HCI, these findings suggest that social support should transition from a supplemental feature to a crucial layer of digital systems. Embedding mechanisms that activate family participation, integrate culturally trusted mediators, and provide alternatives when networks are inactive can ensure that support is both sustainable and culturally resonant.

*Implication 4: Design for cultural pluralism, not cultural generalisation.* Recent evidence highlights that cultural adaptation in DHIs should be viewed as an iterative, resource-intensive process that requires a clear definition of the focal cultural group from the outset [73]. Our corpus reinforces this point: while family involvement strengthened PA engagement among Latina women [57, 64], it acted as a barrier for Korean women who were already overburdened by caregiving duties [22, 92]. Chinese and multiethnic Asian participants, in contrast, emphasised the importance of community trust, contextual usability, and culturally appropriate delivery channels [21, 71].

These divergent patterns highlight that cultural adaptation cannot be assumed or generalised; it should be co-constructed and co-designed with users. Culturally sensitive DHIs require continuous user involvement, careful prioritisation of relevant features, and participatory methods that embed lived experiences into design from the very beginning. Without such approaches, interventions risk reinforcing the very inequities they aim to address.

## 5 Limitations and Future Works

This scoping review followed the PRISMA-ScR guidelines [97] to systematically map and synthesise the landscape of digital PA interventions for culturally and linguistically diverse women. While this methodological approach ensured rigour and transparency, the interpretation of our findings must be contextualised within the inherent constraints of the available literature.

The most significant limitation lies in the geographical and cultural narrowness of the existing evidence base. The majority of included studies were conducted in the United States and focused on a limited number of cultural groups, predominantly Latina women. Consequently, our synthesis and the design paradox it identifies are inherently shaped by the experiences of specific ethnic minorities within a particular national context. The generalisability of our framework to CALD women in other English-speaking countries requires further empirical validation and cannot be assumed.

Furthermore, although our inclusive scope captured a diversity of study designs and digital modalities, the variety of the 18-study corpus limited opportunities for systematic comparison or meta-analysis. Most studies demonstrated efficacy in short-term activity metrics (e.g., daily steps, MVPA) but offered limited evidence concerning sustained engagement or profound cultural adaptation. This observation aligns with our core critique of surface-level localisation: the literature itself frequently lacks the granular design detail necessary to isolate the active ingredients of success beyond a conceptual level.

These limitations define an opportunity for future HCI research. First, there is a critical need to validate and refine the proposed framework across more diverse geographical and cultural contexts to establish its broader utility and boundaries. Second, the field should prioritise longitudinal, theory-driven research that explicitly centres sustained engagement as a primary outcome and provides detailed accounts of how cultural adaptation is translated into concrete interaction design decisions. Finally, rather than pursuing over-generalised solutions, we call for researchers to adopt participatory [39, 84] and co-design approaches [77, 95] that prioritise cultural specificity and contextual depth. Methods such as cultural transcreation workshops [70], scenario-based prototyping [20], and interactive walkthroughs can help translate abstract design principles into situated, culturally resonant strategies. These approaches enable stakeholders to collaboratively map cultural values, social roles, and behavioural motivators into concrete digital interaction patterns [102]. This pathway prioritises depth, long-term sustainability, and cultural resonance over scalability alone, thereby offering a more robust and equitable direction for digital health design.

Looking ahead, this commitment to deep cultural embedding invites a critical examination of emerging technologies, particularly artificial intelligence (AI), which is often promoted for its scalability and personalization capabilities [35, 87]. While AI holds potential for delivering adaptive, highly tailored interventions, these capabilities raise significant ethical considerations [32, 91], including data privacy, algorithmic bias, cultural misclassification, and digital surveillance for marginalised populations. Therefore, any future research in this vein must be coupled with rigorous ethical frameworks and participatory design to ensure equity and trust.

## 6 Conclusion

This scoping review has identified a fundamental design paradox at the heart of digital PA interventions for CALD women: techno-solutionist features tend to ignore complex socio-cultural contexts, while pro-social features often fail in low-activity networks. We have argued that this paradox is a structural outcome of a dominant surface-localisation paradigm that reduces culture to a set of barriers to be overcome. In response, we contributed the Culturally Embedded Interaction Framework, a mid-range design framework comprising five dimensions to guide the creation of socio-technical systems that support long-term cultural pluralism. By translating this framework into actionable design implications, we provide HCI researchers and practitioners with a principled path to move beyond one-size-fits-all accessibility and toward genuinely sustaining, equitable, and culturally resonant design.

## Acknowledgments

Generative AI tools were used at the paragraph level to improve the readability of text originally envisioned and written by the authors. All content, arguments, and conclusions are the authors' own, and the authors take full responsibility for the final manuscript. This work is partially supported by the University of Auckland Faculty of Science Research Development Fund Grant Number 3731533.

# A Appendix A: The final search queries in the syntax of each database

| Database | Query |
|---|---|
| PubMed | (((("immigrant" OR "migrant" OR "foreign-born" OR "newcomer" OR "asylum-seeking" OR "refugee" OR "CALD" OR "ethnic minority" OR "minority women" OR "racial minority" OR "international college student" OR "overseas student" OR "Asian" OR "East Asian" OR "South Asian" OR "Chinese American" OR "Asian immigrant" OR "first-generation immigrant" OR "culturally diverse" OR "ethnic women") AND ("women" OR "female")) AND ("digital intervention" OR "mobile health" OR "mHealth" OR "eHealth" OR "wearable" OR "app" OR "mobile application" OR "technology-based" OR "virtual reality" OR "augmented reality" OR "mixed reality" OR "extended reality" OR "XR" OR "digital health" OR "virtual environment" OR "gamified intervention" OR "digital game" OR "serious game")) AND ("physical activity" OR "exercise" OR "exercise participation" OR "active lifestyle" OR "fitness" OR "fitness engagement" OR "sport participation" OR "exergame" OR "physical activity socialization" OR "exergaming" OR "exercise game" OR "movement-based game" OR "sports game" OR "active play") |
| Scopus | TITLE-ABS-KEY ( ( "immigrant" OR "migrant" OR "refugee" OR "CALD" OR "ethnic minority" OR "culturally diverse" OR "Asian" OR "East Asian" OR "South Asian" OR "Chinese" OR "Asian immigrant" OR "first-generation" ) ) AND TITLE-ABS-KEY ( ( "women" OR "female" ) ) AND TITLE-ABS-KEY ( ( "digital intervention" OR "mobile health" OR "mHealth" OR "eHealth" OR "wearable" OR "app" OR "mobile application" OR "digital health" OR "virtual reality" OR "augmented reality" OR "serious game" OR "exergame" ) ) AND TITLE-ABS-KEY ( ( "physical activity" OR "exercise" OR "fitness" OR "sport" OR "active lifestyle" ) ) |
| Web of Science Core Collection | ((("immigrant" OR "migrant" OR "foreign-born" OR "newcomer" OR "asylum-seeking" OR "refugee" OR "CALD" OR "ethnic minorit*" OR "minorit* women" OR "racial minorit*" OR "international college student" OR "overseas student" OR "Asian" OR "East Asian" OR "South Asian" OR "Chinese American" OR "Asian immigrant" OR "first-generation immigrant" OR "culturally diverse" OR "ethnic women") AND ("women" OR "female")) AND ("digital intervention" OR "mobile health" OR "mHealth" OR "eHealth" OR "wearable" OR "app" OR "mobile application" OR "technology-based" OR "virtual realit*" OR "augmented realit*" OR "mixed realit*" OR "extended realit*" OR "XR" OR "digital health" OR "virtual environment" OR "gamified intervention" OR "digital game" OR "serious game") AND ("physical activity" OR "exercise" OR "exercise participation" OR "active lifestyle" OR "fitness" OR "fitness engagement" OR "sport participation" OR "exergame*" OR "physical activity socialization" OR "exergaming" OR "exercise game" OR "movement-based game" OR "sports game" OR "active play")) |
| ACM Digital Library | [[[All: "immigrant"] OR [All: "migrant"] OR [All: "foreign-born"] OR [All: "newcomer"] OR [All: "asylum-seeking"] OR [All: "refugee"] OR [All: "cald"] OR [All: "ethnic minority"] OR [All: "minority women"] OR [All: "racial minority"] OR [All: "international college student"] OR [All: "overseas student"] OR [All: "asian"] OR [All: "east asian"] OR [All: "south asian"] OR [All: "chinese american"] OR [All: "asian immigrant"] OR [All: "first-generation immigrant"] OR [All: "culturally diverse"] OR [All: "ethnic women"]] AND [[All: "women"] OR [All: "female"]] AND [[All: "digital intervention"] OR [All: "mobile health"] OR [All: "mhealth"] OR [All: "ehealth"] OR [All: "wearable"] OR [All: "app"] OR [All: "mobile application"] OR [All: "technology-based"] OR [All: "virtual reality"] OR [All: "augmented reality"] OR [All: "mixed reality"] OR [All: "extended reality"] OR [All: "xr"] OR [All: "digital health"] OR [All: "virtual environment"] OR [All: "gamified intervention"] OR [All: "digital game"] OR [All: "serious game"]] AND [[All: "physical activity"] OR [All: "exercise"] OR [All: "exercise participation"] OR [All: "active lifestyle"] OR [All: "fitness"] OR [All: "fitness engagement"] OR [All: "sport participation"] OR [All: "exergame"] OR [All: "physical activity socialization"] OR [All: "exergaming"] OR [All: "exercise game"] OR [All: "movement-based game"] OR [All: "sports game"] OR [All: "active play"]] |
| IEEE Xplore Digital Library | ("All Metadata":"immigrant" OR "All Metadata":"migrant" OR "All Metadata":"foreign-born" OR "All Metadata":"newcomer" OR "All Metadata":"asylum-seeking" OR "All Metadata":"refugee" OR "All Metadata":"CALD" OR "All Metadata":"ethnic minority" OR "All Metadata":"minority women" OR "All Metadata":"racial minority" OR "All Metadata":"international college student" OR "All Metadata":"overseas student" OR "All Metadata":"Asian" OR "All Metadata":"East Asian" OR "All Metadata":"South Asian" OR "All Metadata":"Chinese American" OR "All Metadata":"Asian immigrant" OR "All Metadata":"first-generation immigrant" OR "All Metadata":"culturally diverse" OR "All Metadata":"ethnic women") AND ("All Metadata":"women" OR "All Metadata":"female") AND ("All Metadata":"digital intervention" OR "All Metadata":"mobile health" OR "All Metadata":"mHealth" OR "All Metadata":"eHealth" OR "All Metadata":"wearable" OR "All Metadata":"app" OR "All Metadata":"mobile application" OR "All Metadata":"technology-based" OR "All Metadata":"virtual reality" OR "All Metadata":"augmented reality" OR "All Metadata":"mixed reality" OR "All Metadata":"extended reality" OR "All Metadata":"XR" OR "All Metadata":"digital health" OR "All Metadata":"virtual environment" OR "All Metadata":"gamified intervention" OR "All Metadata":"digital game" OR "All Metadata":"serious game") AND ("All Metadata":"physical activity" OR "All Metadata":"exercise" OR "All Metadata":"exercise participation" OR "All Metadata":"active lifestyle" OR "All Metadata":"fitness" OR "All Metadata":"fitness engagement" OR "All Metadata":"sport participation" OR "All Metadata":"exergame" OR "All Metadata":"physical activity socialization" OR "All Metadata":"exergaming" OR "All Metadata":"exercise game" OR "All Metadata":"movement-based game" OR "All Metadata":"sports game" OR "All Metadata":"active play") |

B Appendix B: Characteristics of excluded studies

| No. | Article | Reason for Exclusion | EC Category |
|---|---|---|---|
| 1 | 'I Don't Need a Goal': Attitudes and Practices in Fitness Tracking beyond WEIRD User Groups | Not a PA-promoting digital intervention; examines tracking attitudes/practices only Not focused on CALD women in English-speaking contexts | EC3: Intervention Scope |
| 2 | "It's Great to Exercise Together on Zoom!": Understanding the Practices and Challenges of Live Stream Group Fitness Classes | Not a PA-promoting digital intervention; observational study of practices/challenges rather than an intervention; CALD women not a focal population | EC3: Intervention Scope |
| 3 | Access to health services among culturally and linguistically diverse populations in the Australian universal health care system: issues and challenges | Not a PA-promoting digital intervention; focuses on health-services access rather than physical activity | EC3: Intervention Scope EC2: Interventiion Type |
| 4 | An Electronic Wellness Program to Improve Diet and Exercise in College Students: A Pilot Study | Digital intervention present but not CALD-women specific. | EC1: Population |
| 5 | Apps for IMproving FITness and Increasing Physical Activity Among Young People: The AIMFIT Pragmatic Randomized Controlled Trial | General youth sample; not women-focused; not CALD-focused | EC1: Population |
| 6 | Be Our Guest: Intercultural Heritage Exchange through Augmented Reality (AR) | AR learning/engagement study; no PA outcomes; no women/CALD focal subgroup | EC1: Population EC3: Intervention Scope |
| 7 | Better Supporting Human Aspects in Mobile eHealth Apps: Development and Validation of Enhanced Guidelines | Not a physical-activity–promoting digital intervention; not focused on adult CALD women | EC1: Population EC3: Intervention Scope |
| 8 | Breaking Barriers in Mobile Game Development | Not a physical-activity–promoting digital intervention; not focused on adult CALD women | EC1: Population EC3: Intervention Scope |
| 9 | Designing Leaderboards for Gamification: Perceived Differences Based on User Ranking, Application Domain, and Personality Traits | Generic gamification research; no PA outcomes; no CALD-women focus | EC1: Population EC3: Intervention Scope |
| 10 | Effect of Values and Technology Use on Exercise: Implications for Personalized Behavior Change Interventions | No intervention component; no CALD-women focus | EC1: Population EC3: Intervention Scope |
| 11 | Enhancing Psychosocial Constructs Associated with Technology-Based Physical Activity: A Randomized Trial Among African American Women | Participants are native-born, English-dominant racial minorities with no evidence of migration or refugee background and no non-English household language | EC1: Population |
| 12 | EquityWare: Co-Designing Wearables With And For Low Income Communities In The U.S. | Not a physical-activity–promoting digital intervention; population not specifically adult CALD women | EC1: Population EC3: Intervention Scope |
| 13 | Facilitators, motivations, and barriers to physical activity among Chinese American women | Not a PA-promoting digital intervention; examines facilitators/barriers rather than testing an intervention | EC3: Intervention Scope |
| 14 | From Solo to Social: Exploring the Dynamics of Player Cooperation in a Co-located Cooperative Exergame | Not a PA-promoting digital intervention; exploratory gameplay study rather than an intervention; not focused on adult CALD women | EC1: Population EC3: Intervention Scope |
| 15 | GPTCoach: Towards LLM-Based Physical Activity Coaching | Useful HCI context on LLM coaching; not CALD-women–focused | EC1: Population |
| 16 | Home-based walking intervention for middle-aged migrant women using 360-degree virtual videos and a wearable activity tracker: A mixed-methods pilot study | Study conducted outside Anglophone settings; participants do not meet the review's operational definition of CALD | EC1: Population |
| 17 | Just Dance: The Effects of Exergame Feedback and Controller Use on Physical Activity and Psychological Outcomes | Not a PA-promoting intervention for the target population; observational/experimental gameplay study rather than a designed PA program for adult CALD women | EC1: Population EC3: Intervention Scope |
| 18 | Media format matters: users' perceptions of physical versus digital games | Not a physical-activity–promoting digital intervention; not focused on adult CALD women | EC1: Population EC3: Intervention Scope |
| 19 | Physical and Augmented Reality based Playful Activities for Refresher Training of ASHA Workers in India | Training-focused AR system; no CALD-women target; no PA engagement/adherence outcomes | EC1: Population EC3: Intervention Scope |
| 20 | Ratings and experiences in using a mobile application to increase physical activity among university students: implications for future design | Population not aligned with review focus: not adult CALD wome | EC1: Population |
| 21 | Seek and Reflect: A Mobile Scavenger Hunt to Develop Community Engagement | Not a physical-activity–promoting digital intervention; focuses on community engagement/reflection rather than PA; not focused on adult CALD women | EC1: Population EC3: Intervention Scope |
| 22 | Smartphone apps to improve fitness and increase physical activity among young people: protocol of the Apps for IMproving FITness (AIMFIT) randomized controlled trial | General youth sample; not CALD-women focused; protocol paper without intervention results | EC1: Population EC5: Study Design |
| 23 | Social Virtual Reality as a Mental Health Tool: How People Use VRChat to Support Social Connectedness and Wellbeing | Not a physical-activity–promoting digital intervention; primary focus on mental health/social connectedness rather than PA | EC3: Intervention Scope EC4: Intervention Focus |
| 24 | Socio-Cultural Determinants of Physical Activity among Latin American Immigrant Women in Alberta, Canada | Not a digital, PA-promoting intervention; focuses on barriers/facilitators rather than testing an intervention | EC2: Intervention Type |
| 25 | Socioeconomic Class in Physical Activity Wearables Research and Design | Not a physical-activity–promoting digital intervention; not focused on adult CALD women | EC1: Population EC3: Intervention Scope |
| 26 | StepsBooster-S: A Culturally Tailored Step-Based Persuasive Application for Promoting Physical Activity | Conducted outside Anglophone immigrant contexts; participants are not CALD women living in English-speaking countries; | EC1: Population |
| 27 | That's the way I (dis-)like it! - Effect of Gamer Types and Competitiveness in VR Exergaming | Not a PA-promoting digital intervention for the target population | EC1: Population EC3: Intervention Scope |

| 28 | The Jade Gateway to Exergaming: How Socio-Cultural Factors Shape Exergaming Among East Asian Older Adults | Conducted outside Anglophone immigrant contexts; not a PA-promoting digital intervention; not focused on CALD women | EC1: Population<br>EC3: Intervention Scope |
|---|---|---|---|
| 29 | A Mixed Methods Study on Engagement and Satisfaction with a Digitally-Enhanced Pilot Intervention Among African American and Hispanic Women | Sample does not demonstrate CALD/NESB status: participants were required to read and speak English; no evidence of non-English household language or migration/refugee background | EC1: Population |
| 30 | A Remote Health Coaching, Text-Based Walking Program in Ethnic Minority Primary Care Patients With Overweight and Obesity: Feasibility and Acceptability Pilot Study | Sample required English-speaking participants and "comfortable communicating in English" | EC1: Population |

C Appendix C: Summary of Characteristics of Included Studies

| Summary of Included Studies | | References | | |
|---|---|---|---|---|
| | | Authors/Publication Year/Titles | Relevent Interpretation | |
| | | | Research Location | Population |
| Target Populations | Focus Populations: Women from Asian CALD backgrounds in the United States, with a particular emphasis on Korean American midlife women (n=4) | Choi 2021<br>*A Pilot Study to Promote Active Living among Physically Inactive Korean American Women* | San Francisco Bay Area, California, United States | Physically inactive Korean American women, aged 40–69 years. |
| | | Chee 2016<br>*Practical Issues in Developing a Culturally Tailored Physical Activity Promotion Program for Chinese and Korean American Midlife Women: A Pilot Study* | United States | Chinese and Korean American midlife women (typically aged 40–60 years), in good health, with internet access, and without current or prior medical or cardiovascular conditions. |
| | | Shin 2022<br>*A Technology-Enhanced Physical Activity Intervention: A Feasibility Study* | United States | Midlife Korean American women. |
| | | Nguyen 2024<br>*An App-Based Physical Activity Intervention in Community-Dwelling Chinese-, Tagalog-, and Vietnamese-Speaking Americans* | San Francisco Bay Area, California, United States | Nineteen physically underactive Asian American adults (aged 25–70 years), of whom 58% (n = 11) were female; participants spoke Chinese (Mandarin/Cantonese), Tagalog, or Vietnamese. |
| | Focus populations: Women from CALD backgrounds in the United States, with a predominant focus on Latina populations, particularly Spanish-speaking women (n=14) | Rowland 2022<br>*Feasibility, Usability and Acceptability of a mHealth Intervention to Reduce Cardiovascular Risk in Rural Hispanic Adults* | A rural agricultural region in California, United States | Seventy Hispanic/Latino adults (aged 19–65 years), both men and women, with the majority being female. |
| | | Figueroa 2021<br>*Conversational Physical Activity Coaches for Spanish- and English-Speaking Women: A User Design Study* | Northern California, United States | Twenty-six low-income women (aged 18–65 years) residing in the United States, including 16 Spanish-speaking and 10 English-speaking participants, all physically inactive and from underserved communities. |
| | | Bender 2016<br>*Designing a Culturally Appropriate Visually Enhanced Low-Text Mobile Health App Promoting Physical Activity for Latinos: A Qualitative Study* | California, United States | Spanish-speaking Latino adults. |
| | | Collins 2019<br>*Efficacy of a multi-component intervention to promote physical activity among Latino adults: A randomized controlled trial* | Wichita, Kansas, United States | Latino adults. |
| | | Benitez 2021<br>*Design and rationale for a randomized trial of a theory- and technology- enhanced physical activity intervention for Latinas: The Seamos Activas II study* | San Diego County, California, United States | Latina women (aged 18–65 years), physically underactive (<60 minutes of MVPA per week). |
| | | Marcus 2022<br>*Physical activity outcomes from a randomized trial of a theory- and technology-enhanced intervention for Latinas: the Seamos Activas II study* | San Diego County, California, United States | Underactive Latina women (aged 18–65 years), Spanish-reading; approximately 89% were of Mexican/Mexican-American background. |
| | | Marcus 2021<br>*Long-term physical activity outcomes in the Seamos Activas II trial* | San Diego County, California, United States | Underactive Latina women (aged 18–65 years), Spanish-speaking. |
| | | Mendoza-Vasconez 2022<br>*Regular and App-enhanced Maintenance of Physical Activity among Latinas: A Feasibility Study* | San Diego County, California, United States | Latina women. |
| | | Benitez 2016<br>*Using web-based technology to promote physical activity in Latinas: Results of the Muévete Alabama pilot study* | San Diego County, California, United States | Spanish-speaking Latina women who were physically underactive. |
| | | Marcus 2015<br>*Using interactive Internet technology to promote physical activity in Latinas: Rationale, design, and baseline findings of Pasos Hacia La Salud* | San Diego County, California, United States | Spanish-speaking Latina women who were physically underactive. |
| | | Marcus 2016<br>*Pasos Hacia La Salud: a randomized controlled trial of an internet-delivered physical activity intervention for Latinas* | San Diego County, California, United States | Underactive Spanish-speaking Latina women, primarily of Mexican/Mexican-American background. |
| | | Linke 2019<br>*Association Between Physical Activity Intervention Website Use and Physical Activity Levels Among Spanish-Speaking Latinas* | San Diego County, California, United States | Spanish-speaking Latina women (aged 18–65 years), physically inactive (<60 minutes of MVPA per week). |
| | | Bohlen 2024<br>*Six-Month Outcomes of a Theory- and Technology-Enhanced Physical Activity Intervention for Latina Women (Pasos Hacia La Salud II): Randomized Controlled Trial* | Providence, Rhode Island, United States | Spanish-speaking Latina women. |
| | | Ash 2025<br>*Pasos Hacia La Salud II: A Superiority RCT Utilizing Technology to Promote Physical Activity in Latinas* | Providence, Rhode Island, United States | Spanish-speaking Latina women (aged 18–65 years) who were physically inactive (<150 minutes of MVPA per week). |
| Study Design | Randomized Controlled Trials (RCTs) | Choi 2021<br>*A Pilot Study to Promote Active Living among Physically Inactive Korean American Women* | In a pilot controlled trial, 40 women were randomized to WALK-regular or WALK-plus. | |
| | | Collins 2019<br>*Efficacy of a multi-component intervention to promote physical activity among Latino adults: A randomized controlled trial* | This study conducted a one-year, NIH funded, randomized, investigator-blinded clinical trial in a cohort of Latino adults at risk for CVD. | |
| | | Benitez 2020<br>*Design and rationale for a randomized trial of a theory- and technology- enhanced physical activity intervention for Latinas: The Seamos Activas II study* | Seamos Activas II is a 6-month randomized controlled trial comparing two PA intervention conditions in Latina women | |
| | | Marcus 2022<br>*Physical activity outcomes from a randomized trial of a theory- and technology-enhanced intervention for Latinas: the Seamos Activas II study* | Seamos Activas II was a 6-month randomized controlled trial | |
| | | Marcus 2021<br>*Long-term physical activity outcomes in the Seamos Activas II trial* | Seamos Activas II was a 6-month randomized controlled trial with a maintenance phase from 6 to 12 months that compared two PA interventions for Latina women: | |
| | | Marcus 2016<br>*Pasos Hacia La Salud: a randomized controlled trial of an internet-delivered physical activity intervention for Latinas* | The Pasos Hacia La Salud study (N = 205) was a randomized controlled trial of a 6-month | |
| | | Marcus 2015<br>*Using interactive Internet technology to promote physical activity in Latinas: Rationale, design, and baseline findings of Pasos Hacia La Salud* | The Pasos Hacia La Salud study (N=218) is a randomized controlled trial of a 6-month | |
| | | Linke 2019<br>*Association Between Physical Activity Intervention Website Use and Physical Activity Levels Among Spanish-Speaking Latinas* | The Pasos Hacia La Salud study was a randomized controlled trial of an internet-based PA intervention versus a wellness contact control in Latinas | |
| | | Bohlen 2024<br>*Six-Month Outcomes of a Theory- and Technology-Enhanced Physical Activity Intervention for Latina Women (Pasos Hacia La Salud II): Randomized Controlled Trial* | This 6-month randomized controlled trial | |
| | | Ash 2024<br>*Pasos Hacia La Salud II: A Superiority RCT Utilizing Technology to Promote Physical Activity in Latinas* | The research team conducted a superiority randomized controlled trial (RCT) | |
| | Feasibility study | Shin 2022<br>*A Technology-Enhanced Physical Activity Intervention: A Feasibility Study* | A Feasibility Study: Using a single-group, pre-posttest design, this study evaluated feasibility and acceptability of a technology-enhanced physical activity intervention | |
| | | Nguyen 2024<br>*An App-Based Physical Activity Intervention in Community-Dwelling Chinese-, Tagalog-, and Vietnamese-Speaking Americans* | This was a single-arm, 5-week (1-week run-in period and 4-week intervention period), pilot interventional study | |
| | | Benitez 2016<br>*Using web-based technology to promote physical activity in Latinas: Results of the Muévete Alabama pilot study* | In this pilot study, Muévete Alabama, a single-arm pre-posttest design was utilized to assess the feasibility and acceptability of this Spanish language and individually tailored physical activity website for Latinas. | |
| | | Rowland 2022<br>*Feasibility, Usability and Acceptability of a mHealth Intervention to Reduce Cardiovascular Risk in Rural Hispanic Adults* | A descriptive study using quantitative and qualitative methods was used to evaluate the feasibility, usability, and acceptability of delivering a 12-week mHealth self-management intervention. | |
| | mixed-methods research design | Chee 2016<br>*Practical Issues in Developing a Culturally Tailored Physical Activity Promotion Program for Chinese and Korean American Midlife Women: A Pilot Study* | Pilot mixed-methods study with usability testing, expert review, and a preliminary randomized controlled trial (RCT) among Chinese and Korean American midlife women. | |
| | | Mendoza-Vasconez 2022<br>*Regular and App-enhanced Maintenance of Physical Activity among Latinas: A Feasibility Study* | This study used a convergent mixed-methods research design, incorporating quantitative and qualitative methods in parallel. | |

| | Summary of Included Studies | References | |
|---|---|---|---|
| | | Authors/Publication Year/Titles | Relevent Interpretation |
| | **Qualitative Studies** | Bender 2016<br>Designing a Culturally Appropriate Visually Enhanced Low-Text Mobile Health App Promoting Physical Activity for Latinos: A Qualitative Study | A qualitative exploratory study using focus groups and in-depth interviews |
| **Goals for improve physical activity** | **Chronic Disease Prevention and Risk Reduction** | Choi 2021<br>A Pilot Study to Promote Active Living among Physically Inactive Korean American Women | Reduce the risk of Metabolic syndrome (MetS) |
| | | Nguyen 2024<br>An App-Based Physical Activity Intervention in Community-Dwelling Chinese-, Tagalog-, and Vietnamese-Speaking Americans | Reduce the risk of Metabolic syndrome (MetS), prevent and moderate chronic diseases |
| | | Rowland 2022<br>Feasibility, Usability and Acceptability of a mHealth Intervention to Reduce Cardiovascular Risk in Rural Hispanic Adults | support the self-management of chronic health conditions like hypertension and diabetes |
| | | Figueroa 2021<br>Conversational Physical Activity Coaches for Spanish- and English-Speaking Women: A User Design Study | Improve physical activity and other health related aspects |
| | | Chee 2016<br>Practical Issues in Developing a Culturally Tailored Physical Activity Promotion Program for Chinese and Korean American Midlife Women: A Pilot Study | Prevent the progression of metabolic syndrome and related chronic diseases by fostering sustainable physical activity habits among midlife women, a population at elevated risk due to high inactivity levels. |
| | | Bender 2016<br>Designing a Culturally Appropriate Visually Enhanced Low-Text Mobile Health App Promoting Physical Activity for Latinos: A Qualitative Study | Improve leisure-time physical activity among Latinos to reduce obesity and obesity-related diseases, while supporting health literacy and healthier lifestyle behaviors such as increased PA and reduced sedentary behavior. |
| | | Shin 2022<br>A Technology-Enhanced Physical Activity Intervention: A Feasibility Study | Reduce cardiometabolic disease (CMD) risks among midlife Korean American adults by promoting physical activity (PA), which can prevent one in ten premature deaths, improve healthy aging, and lower associated healthcare costs. |
| | | Collins 2019<br>Efficacy of a multi-component intervention to promote physical activity among Latino adults: A randomized controlled trial | Address cardiovascular disease (CVD) risks among Latinos by targeting highly prevalent factors such as type 2 diabetes, hyperlipidemia, and physical inactivity, which together exacerbate vulnerability to poor cardiometabolic outcomes. |
| | | Benitez 2020<br>Design and rationale for a randomized trial of a theory- and technology- enhanced physical activity intervention for Latinas: The Seamos Activas II study | Emphasize the importance of maintaining healthy PA levels among Latinas in the U.S., who face disproportionately higher risks—being 70% more likely to develop diabetes and 20% more likely to be obese than non-Latina White women—yet only 44% meet national PA guidelines compared with 55% of non-Latina White women |
| | | Marcus 2022<br>Physical activity outcomes from a randomized trial of a theory- and technology-enhanced intervention for Latinas: the Seamos Activas II study | Address the urgent need to improve PA among Latina women, who report disproportionately high and increasing prevalence of obesity and diabetes, conditions strongly associated with inactivity, placing them at elevated risk for long-term cardiometabolic disease and health inequities. |
| | | Marcus 2021<br>Long-term physical activity outcomes in the Seamos Activas II trial | Address the disproportionate prevalence of lifestyle-related health conditions among Latinas in the U.S., including diabetes, stroke, and obesity, by promoting sustained physical activity engagement as a key preventive strategy. |
| | | Mendoza-Vasconez 2022<br>Regular and App-enhanced Maintenance of Physical Activity among Latinas: A Feasibility Study | Address disproportionately low activity rates, improve long-term health outcomes |
| | | Benitez 2016<br>Using web-based technology to promote physical activity in Latinas: Results of the Muévete Alabama pilot study | Reduce the risk of type 2 diabetes and obesity through sustained engagement in physical activity. |
| | | Benitez 2016<br>Using interactive Internet technology to promote physical activity in Latinas: Rationale, design, and baseline findings of Pasos Hacia La Salud | Prevent the onset of multiple chronic conditions, including cardiovascular disease, type 2 diabetes, and hypertension, by sustaining regular physical activity. |
| | | Marcus 2016<br>Pasos Hacia La Salud: a randomized controlled trial of an internet-delivered physical activity intervention for Latinas | Prevent the onset of multiple chronic conditions, including cardiovascular disease, type 2 diabetes, and hypertension, by sustaining regular physical activity. |
| | | Linke 2019<br>Association Between Physical Activity Intervention Website Use and Physical Activity Levels Among Spanish-Speaking Latinas | Prevent the onset of multiple chronic conditions, including cardiovascular disease, type 2 diabetes, and hypertension, by sustaining regular physical activity. |
| | | Bohlen 2024<br>Six-Month Outcomes of a Theory- and Technology-Enhanced Physical Activity Intervention for Latina Women (Pasos Hacia La Salud II): Randomized Controlled Trial | Reduce the likelihood of developing type 2 diabetes and hypertension, conditions projected to affect more than half of Latina women. |
| | **Enhance psychological well-being by improving mood** | Ash 2024<br>Pasos Hacia La Salud II: A Superiority RCT Utilizing Technology to Promote Physical Activity in Latinas | Emphasize the importance of aerobic activity for U.S. Latinas, who are less physically active than non-Hispanic White women. Regular physical activity can prevent and control chronic health conditions, support healthy weight management, and enhance psychological well-being by improving mood. |

D  Appendix D: Thematic Synthesis of Included Studies.pdf

| Thematic Synthesis | No | Summary of Included Studies | References | |
|---|---|---|---|---|
| | | | Authors/Publication Year/Titles | Relevent Interpretation |
| 1. Surface level-cultural adpatation | | Interface-Centered Adaptations for Digital Accessibility (n=9) | Collins 2019<br>*Efficacy of a multi-component intervention to promote physical activity among Latino adults: A randomized controlled trial* | Surface-level cultural adaptation<br>This intervention is not culturally tailored in a deep, structural, or culturally grounded way. It uses linguistic translation and message tailoring by stage of change, but it does not redesign the digital intervention based on Latino cultural values, beliefs, social norms, or lived experiences.<br>Content Tailoring Based on Readiness: Messages were tailored to the participant's PACE score (stage of readiness). This is a behavioral theory-based adaptation, not necessarily a cultural one.<br>Motivational Interviewing (MI): MI is a general counseling approach that is client-centered and can be culturally flexible, but the study does not describe how MI was specifically adapted for Latino cultural values (e.g., familismo, respeto, personalismo). |
| | | | Mendoza-Vasconez 2022<br>*Regular and App-enhanced Maintenance of Physical Activity among Latinas: A Feasibility Study* | Tested Spanish-language app features (goal setting, reminders, tracking) to support maintenance; addressed usability and digital literacy but did not incorporate deeper cultural or emotional tailoring. |
| | | | Choi 2021<br>*A Pilot Study to Promote Active Living among Physically Inactive Korean American Women* | Surface-Level<br>The WALK program for Korean American women involved surface-level cultural and linguistic adaptation—Korean-language materials, recruitment through Korean community networks, use of popular Korean messaging platforms (KakaoTalk, NAVER BAND), and input from a Korean Community Advisory Board—but did not substantially redesign behavior-change strategies or digital features based on Korean American cultural values, gender norms, or immigration-related experiences. |
| | | | Rowland 2022<br>*Feasibility, Usability and Acceptability of a mHealth Intervention to Reduce Cardiovascular Risk in Rural Hispanic Adults* | Surface-Level<br>Rowland et al. incorporated community actors through a community advisory board that guided recruitment and ensured the cultural relevance of intervention SMS content; however, cultural adaptation remained largely surface-level, as the core MyFitnessPal interaction and feature set were not redesigned to reflect culturally specific practices or gendered relational contexts. |
| | | | Figueroa 2021<br>*Conversational Physical Activity Coaches for Spanish- and English-Speaking Women: A User Design Study* | Surface-Level<br>This user-design study developed a bilingual PA chatbot prototype and used co-creation with low-income English- and Spanish-speaking women to explore preferences around language, privacy, digital literacy, and family inclusion. However, the implemented conversational flows remained generic (values, goals, barriers) and were not yet structurally redesigned around Latina cultural values or social roles. We therefore classified the current intervention as surface-level cultural adaptation, with a strong culturally informed design process aimed at enabling deeper adaptation in future versions.<br>• Features that directly encode family involvement (e.g., messaging kids/partners, shared goals) - it's a future wish from participants, not implemented yet.<br>• This research provides great user-centred and culturally aware processes, but the article mostly reports preferences and design recommendations, not a fully implemented, culture-embedded chatbot. |
| | | | Nguyen 2024<br>*An App-Based Physical Activity Intervention in Community-Dwelling Chinese-, Tagalog-, and Vietnamese-Speaking Americans* | Surface-Level<br>The research was primarily surface-level, consisting of language translation, simplified text, and language-concordant staff facilitation. The app's behavioural strategies, user interaction logic, goal progression, and digital features were directly inherited from the original English-language program without structural modification to reflect cultural values, norms, or motivational patterns of Chinese-, Vietnamese-, or Filipino-speaking Asian Americans.<br>• Translated the app into Traditional Chinese, Tagalog, and Vietnamese<br>• Used bilingual / bicultural staff to check reading level (fourth-grade level)<br>• Conducted forward translation + review<br>• Conducted pretesting with 4 bicultural/bilingual community members<br>• Made language-equivalence and clarity adjustments<br>• Provided in-language text messages, reminders |
| | | | Chee 2016<br>*Practical Issues in Developing a Culturally Tailored Physical Activity Promotion Program for Chinese and Korean American Midlife Women: A Pilot Study* | Surface-Level<br>Although this web-based PA program for Chinese and Korean American midlife women used multiple languages (English/Chinese/Korean), bilingual staff, and culturally sensitive recruitment/communication strategies, the underlying intervention structure (forums, education modules, email coaching) and digital features were not fundamentally redesigned based on Chinese/Korean American cultural values or lived experiences; therefore we classified it as surface-level rather than deep cultural adaptation. |
| | | | Bender 2016<br>*Designing a Culturally Appropriate Visually Enhanced Low-Text Mobile Health App Promoting Physical Activity for Latinos: A Qualitative Study"* | Surface-level cultural adaptation<br>This study is a formative qualitative exploration that identifies culturally relevant imagery for a future Latino mHealth app. However, it does not yet translate cultural insights into the intervention architecture, behaviour-change mechanisms, or digital interaction design. Therefore, although culturally sensitive at the visual level, it should be classified as surface-level cultural adaptation. |
| | | | Shin 2022<br>*A Technology-Enhanced Physical Activity Intervention: A Feasibility Study* | Surface-level cultural adaptation<br>"Cultural tailoring" here is operationalised mainly as Korean-language materials and bilingual delivery, plus Fitbit self-monitoring + SMS reminders. The digital mechanism itself is not culturally redesigned. |
| 2. Deep social-cultural adpatation | | Beyond Translation:<br>(1) Cultural and Emotional Alignment;<br>(2) Community Involvement through Cultural Brokers;<br>(3) Culturally Grounded Behaviour Change Design. | Benitez 2020<br>*Design and rationale for a randomized trial of a theory- and technology-enhanced physical activity intervention for Latinas: The Seamos Activas II study"*<br>**Seamos Activas II - 1** | Deep Cultural Adaptation.<br>The study integrates SCT and TTM not just theoretically, but functionally, using tailored materials based on participants' self-efficacy, social support needs, stage of change, and preferences. They started with print-based materials, then added interactive texts, goal-setting calls, and location-based resources. The study explicitly states that the intervention was developed following extensive qualitative research with Latinas to identify culturally relevant barriers to physical activity. The intervention uses the SCT and the TTM. While the theories are universal, their application here is tailored to key social mechanisms in the Latino community, such as leveraging family social support for PA. |
| | | | Marcus 2022<br>*Physical activity outcomes from a randomized trial of a theory- and technology-enhanced intervention for Latinas: the Seamos Activas II study*<br>**Seamos Activas II - 2** | Deep Cultural Adaptation.<br>Seamos Activas II builds on the deeply culturally and linguistically adapted Seamos Saludables program, originally developed through formative work with Latinas and tailored to Latina-specific barriers such as caregiving roles and neighborhood safety, while adding technology-enhanced components (SCT-based text messages and local PA maps). We therefore classified the overall intervention as deep-level cultural adaptation, with the new digital enhancements providing theory-based support rather than additional deep cultural redesign.<br>• Program level: deep cultural adaptation (Spanish, Latina-specific barriers, formative work, culturally tailored print content).<br>• New digital pieces (texts, mapping): mainly theory-driven, context-sensitive, but not strongly re-engineered around culture, more surface-to-moderate on the digital side. |
| | | | Marcus 2021<br>*Long-term physical activity outcomes in the Seamos Activas II trial*<br>**Seamos Activas II -3 - Maintenance phrase** | Deep Cultural Adaptation.<br>1. Adaptation Through Formative Research (Deep Level Foundation)<br>Both the Original and Enhanced versions of the PA intervention were programs that were " culturally adapted for Latinas using formative research".<br>2. Deep Adaptation in the Digital Design and Technology Enhancements<br>Mitigation of Environmental Barriers: The web-based or remote delivery format (used in prior studies like Muévete Alabama and Pasos Hacia La Salud) was a deep structural choice made specifically to "overcome barriers common in face-to-face interventions" such as lack of transportation, childcare, and family responsibilities.<br>3. Targeting Culturally Relevant Constructs: The technology was used to implement theory in a culturally informed way. The Enhanced Intervention targeted specific Social Cognitive Theory (SCT) constructs (social support, enjoyment, and outcome expectancies), which were identified through feedback from Latinas as critical for achieving the highest PA increases. The text messages were specifically designed to support these constructs through goal setting and self-monitoring. |
| | | | Benitez 2016<br>*Using web-based technology to promote physical activity in Latinas: Results of the Muévete Alabama pilot study*<br>**Pilot study of Pasos Hacia La Salud study - 1** | Deep Cultural Adaptation.<br>The intervention was culturally and linguistically adapted. This included translation/back translation of all materials. Extensive formative research was conducted, including six focus groups on PA barriers and a subsequent round of focus groups (n=19) specifically addressing Internet usage among Latinas to inform intervention refinement. The web-based format itself was a deep adaptation designed to overcome barriers like lack of transportation, childcare duties, and fear of immigration authorities prevalent in this population, particularly in Alabama. |
| | | | Marcus 2015<br>*Using interactive Internet technology to promote physical activity in Latinas: Rationale, design, and baseline findings of Pasos Hacia La Salud*<br>**Pasos Hacia La Salud - 2** | Deep Cultural Adaptation.<br>The program was "culturally and linguistically adapted" for the target population through "extensive formative research"". This research included focus groups with Latinas to understand their Internet use behaviours and PA needs. Content was designed to address gender-specific roles and responsibilities (e.g., caregiving and home management) that interfere with PA. The intervention was delivered in Spanish based on participant preference. |
| | | | Marcus 2016<br>*Pasos Hacia La Salud: a randomized controlled trial of an internet-delivered physical activity intervention for Latinas*<br>**Pasos Hacia La Salud -3** | Deep Cultural Adaptation.<br>Recurring Theme: Contextual Adaptation via Formative Research: The deep adaptation was driven by formative research and the need to overcome specific contextual barriers identified in the target population. The development team conducted a series of focus groups with Latinas specifically regarding their Internet use behaviours (why, when, and how often they use the web, and types of sites they visit). This information was used to build the web-based version of the intervention. The digital, home-based nature of the intervention itself serves as a deep adaptation because it directly mitigates key barriers frequently cited by Latinas for participating in traditional face-to-face interventions, such as limited transportation and childcare duties. |
| | | | Linke 2019<br>*Association Between Physical Activity Intervention Website Use and Physical Activity Levels Among Spanish-Speaking Latinas*<br>**Pasos Hacia La Salud II -1** | Deep Cultural Adaptation.<br>Tailoring based on Psychosocial Factors (Theoretical Adaptation): The core of the intervention involved tailoring and personalization of messages based on theoretical constructs derived from the Transtheoretical Model (TTM) and Social Cognitive Theory (SCT). This tailoring addressed psychosocial and environmental factors affecting PA. For example, higher website engagement was positively associated with greater self-efficacy and social support (specifically, the rewards and punishment subscale) at 6 months. The success of tailoring suggests that the theoretically derived messages were culturally and individually relevant." |

| Thematic Synthesis | No | Summary of Included Studies | References | |
|---|---|---|---|---|
| | | | Authors/Publication Year/Titles | Relevant Interpretation |
| | | | Bohlen 2024<br>*Six-Month Outcomes of a Theory- and Technology-Enhanced Physical Activity Intervention for Latina Women (Pasos Hacia La Salud II): Randomized Controlled Trial*<br>**Pasos Hacia La Salud II -2** | Deep Cultural Adaptation.<br>Pasos Hacia La Salud II is built on a Spanish-language intervention developed through formative research with Latina women, with materials that address Latina-specific barriers (e.g., caregiving, safety, time constraints). At the program level, it represents deep cultural adaptation.<br>The digital delivery (web platform, text messaging, and gamified engagement features) is culturally informed but not fully re-engineered around cultural logics. The website and text messages are Spanish-language and incorporate Latina-relevant examples (e.g., dancing) and community contexts, yet the underlying interaction structure. |
| | | | Ash 2024<br>*Pasos Hacia La Salud II: A Superiority RCT Utilizing Technology to Promote Physical Activity in Latinas*<br>**Pasos Hacia La Salud II -3** | Deep Cultural Adaptation.<br>The website includes Spanish-language, stage-matched manuals and tailored feedback and tips for overcoming barriers, with examples such as childcare responsibilities, neighbourhood safety, and lack of transport—issues highlighted in earlier Latina-focused work.<br>The intervention is highly tailored (e.g., to TTM stage, to self-efficacy), but tailoring is a universal behavioral strategy. The content of the tips, text messages, and manuals, while in Spanish, is not described as being specifically designed around core Latino cultural values like familismo. |
| | | *Cultural and Temporal Misalignments* | Chee 2016<br>*Practical Issues in Developing a Culturally Tailored Physical Activity Promotion Program for Chinese and Korean American Midlife Women: A Pilot Study* | Explicit cultural conflict surfaced around the labels "midlife" and "depression" (stigma), plus divergent lay definitions of physical activity; translation asymmetries further undermined cultural fit. |
| | | | Rowland 2022<br>*Feasibility, Usability and Acceptability of a mHealth Intervention to Reduce Cardiovascular Risk in Rural Hispanic Adults* | MyFitnessPal lacked common Latino foods and required burdensome portion entry; device/app syncing failures further undermined motivation. |
| | | | Figueroa 2021<br>*Conversational Physical Activity Coaches for Spanish- and English-Speaking Women: A User Design Study* | cultural mismatch in tone/phrasing and impersonal chatbot responses; participants, especially monolingual Spanish speakers wanted family-inclusive features and stronger personalisation. |
| **Configurations of Technology Enhanced Delivery** | 1 | *Digital-Only Programs* | Nguyen 2024<br>*An App-Based Physical Activity Intervention in Community-Dwelling Chinese-, Tagalog-, and Vietnamese-Speaking Americans* | Digital only<br>Mobile health (mHealth) intervention using a smartphone app + Fitbit + SMS reminders |
| | | | Bender 2016<br>*Designing a Culturally Appropriate Visually Enhanced Low-Text Mobile Health App Promoting Physical Activity for Latinos: A Qualitative Study* | Visual, low-text mobile app prototype for health promotion |
| | | | Benitez 2016<br>*Using web-based technology to promote physical activity in Latinas: Results of the Muévete Alabama pilot study* | Digital-Only Interventions<br>Internet-based, theory-driven, culturally adapted intervention |
| | | | Marcus 2015<br>*Using interactive Internet technology to promote physical activity in Latinas: Rationale, design, and baseline findings of Pasos Hacia La Salud* | Digital-Only Interventions<br>Internet-based, theory-driven, individually tailored, culturally adapted program |
| | | | Marcus 2016<br>*Pasos Hacia La Salud: a randomized controlled trial of an internet-delivered physical activity intervention for Latinas* | Digital-Only Interventions<br>Intervention group: Tailored Internet program (Spanish-language), weekly updates, individualised content (tips, feedback, PA strategies) |
| | | | Figueroa 2021<br>*Conversational Physical Activity Coaches for Spanish- and English-Speaking Women: A User Design Study* | Hybrid / Mult-component Intervention: Digital + Virtual Social Support (non-human)<br>User-centred, SMS-based conversational physical activity intervention using a chatbot designed to coach and motivate participants through personalised, culturally tailored messages. |
| | | | Choi 2021<br>*A Pilot Study to Promote Active Living among Physically Inactive Korean American Women* | WALK-regular : Digital only (for tracking and self-monitoring) - Fitbit, Fitbit app, weekly self-monitoring sheets |
| | | | Linke 2019<br>*Association Between Physical Activity Intervention Website Use and Physical Activity Levels Among Spanish-Speaking Latinas* | Digital-Only Interventions<br>Intervention: Individually tailored, culturally adapted Internet PA program |
| | 2 | *Hybrid Intervention: Digital Interventions with Remote Social Support* | Marcus 2021<br>*Long-term physical activity outcomes in the Seamos Activas II trial* | Automated SMS (tips, weekly goal-setting & reporting with auto-feedback); individualized PA-location mapping; remote phone check-ins |
| | | | Bohlen 2024<br>*Six-Month Outcomes of a Theory- and Technology-Enhanced Physical Activity Intervention for Latina Women (Pasos Hacia La Salud II): Randomized Controlled Trial* | Website + text messages, phone calls, gamification features ( points system, badeges, medals, progress tracking, incentives). |
| | | | Chee 2016<br>*Practical Issues in Developing a Culturally Tailored Physical Activity Promotion Program for Chinese and Korean American Midlife Women: A Pilot Study* | Web-based culturally tailored physical activity intervention including:   Educational modules (on physical activity, menopause, and depression), Weekly discussion forums one-on-one email coaching. |
| | | | Choi 2021<br>*A Pilot Study to Promote Active Living among Physically Inactive Korean American Women* | WALK-plus: Digital with remote support - Fitbit, Fitbit app, weekly self-monitoring sheets, plus remote feedback and reinforcement via NAVER BAND online social community. |
| | | | Collins 2019<br>*Efficacy of a multi-component intervention to promote physical activity among Latino adults: A randomized controlled trial* | Hybrid / Mult-component Intervention: Digital + Remote Social Support<br>Fitbit devices, smartphone apps, and SMS text messages, with motivational interviewing phone calls. |
| | | | Ash 2025<br>*Pasos Hacia La Salud II: A Superiority RCT Utilizing Technology to Promote Physical Activity in Latinas* | Enhanced Web + Text + Coaching Intervention<br>weekly text messages, extra phone calls, gamification (points, medals, incentives), audio manuals, and more social support via the discussion board. |
| | 3 | *Hybrid Intervention: Digital Interventions with In-Person Social Support* | Rowland 2022<br>*Feasibility, Usability and Acceptability of a mHealth Intervention to Reduce Cardiovascular Risk in Rural Hispanic Adults* | Hybrid / Mult-component Intervention: Digital + In-Person Human Support<br>Mobile health intervention using the MyFitnessPal app + Bluetooth-enabled smart scale + SMS reminders + Tech support visit. |
| | | | Shin 2022<br>*A Technology-Enhanced Physical Activity Intervention: A Feasibility Study* | Hybrid / Mult-component Intervention: Digital + In-Person Human Support<br>Participants were given a Fitbit Charge 2 and instructed to wear it on a nondominant hand throughout the entire 12-week intervention period  + Weekly reminder text messages were sent to participants during the entire 12-week intervention period + Bi-weekly, face-to-face, individual or small group (two to three people) education sessions were conducted during the first 4 weeks of the intervention +Monthly small group (two to three people) free style walking sessions followed during the next 8 weeks of the intervention. |
| | 4 | *Hybrid Intervention: Digital Interventions with Remote and In-Person Social Support* | Mendoza-Vasconez 2022<br>*Regular and App-enhanced Maintenance of Physical Activity among Latinas: A Feasibility Study* | App-enhanced maintenance group: at the start of the 3-month maintenance, they are taught in a ~1-hour session how to use two commercial smartphone apps (Fitbit, Nike Training Club), and staff help them download and personalise the apps (goals, etc.). They then use the apps on their own for 3 months. A scheduled phone call at 1 week allows questions about using the apps; no further structured coaching. |
| | | | Marcus 2022<br>*Physical activity outcomes from a randomized trial of a theory- and technology-enhanced intervention for Latinas: the Seamos Activas II study* | Digital components + Same baseline and 6-month goal-setting sessions with staff as the Original arm. + Same month-1 telephone call plus additional telephone calls at months 2 and 3 to encourage use of the text messages and address barriers |
| | | | Benitez 2020<br>*Design and rationale for a randomized trial of a theory- and technology- enhanced physical activity intervention for Latinas: The Seamos Activas II study* | Hybrid / Mult-component Intervention: Digital + Virtual Human Support<br>The core enhancement was the use of interactive text-messaging to increase interactivity and accountability + Yes. Original 1-month call plus additional phone calls at 2, 3, and 9 months to support engagement, especially with the texting component.  + Goal-setting session with trained staff at baseline (start of intervention) and 6-month visit. |
| **Role of Social Support** | 1 | *Virtual Social Support* | Choi 2021<br>*A Pilot Study to Promote Active Living among Physically Inactive Korean American Women* | The study's intervention leveraged culturally familiar virtual platforms (KakaoTalk, Naver BAND) to build online peer groups. These facilitated role modelling and accountability when participants engaged actively, but women embedded in low-activity offline networks reported limited reinforcement. |
| | | | Figueroa 2021<br>*Conversational Physical Activity Coaches for Spanish- and English-Speaking Women: A User Design Study* | A bilingual conversational agent (chatbot) provided personalised encouragement and PA coaching via SMS. |
| | | | Chee 2016<br>*Practical Issues in Developing a Culturally Tailored Physical Activity Promotion Program for Chinese and Korean American Midlife Women: A Pilot Study* | The program embedded social/peer support via Facebook to prompt discussion, weekly goal-setting, and emotional support. Yet the forum architecture (latest-post surfacing) fragmented threads across staggered cohorts, impeding cumulative conversations. |
| | | | Benitez 2020<br>*Design and rationale for a randomized trial of a theory- and technology- enhanced physical activity intervention for Latinas: The Seamos Activas II study* | The study combined interactive SMS, personalised feedback, and remote counselling with culturally tailored print materials. This approach strengthened self-efficacy and accountability, but baseline data showed persistently low levels of friend support. |
| | 2 | *In person Soical Support* | Rowland 2022<br>*Feasibility, Usability and Acceptability of a mHealth Intervention to Reduce Cardiovascular Risk in Rural Hispanic Adults* | The study paired MyFitnessPal and smart devices with in-person technical support visits. These human touchpoints were critical in overcoming digital literacy barriers, enhancing trust. However, effects were often temporary, as adherence declined once in-person support was withdrawn. |
| | | | Chee 2016<br>*Practical Issues in Developing a Culturally Tailored Physical Activity Promotion Program for Chinese and Korean American Midlife Women: A Pilot Study* | Response to individual coaching was low; participants felt judged/pressured by progress checks and avoided one-on-one contact, while overall participation burden (weekly postings, multiple modules) further dampened engagement. |
| | | | Shin 2022<br>*A Technology-Enhanced Physical Activity Intervention: A Feasibility Study* | This study integrated face-to-face group sessions with digital reminders, which fostered belonging and improved PA and cardiometabolic outcomes. However, perceived family and friend support remained largely unchanged. |
| | | | Rowland 2022<br>*Feasibility, Usability and Acceptability of a mHealth Intervention to Reduce Cardiovascular Risk in Rural Hispanic Adults* | |
| | | | Collins 2019<br>*Efficacy of a multi-component intervention to promote physical activity among Latino adults: A randomized controlled trial* | |
| | | | Benitez 2020<br>*Design and rationale for a randomized trial of a theory- and technology- enhanced physical activity intervention for Latinas: The Seamos Activas II study* | |

| Thematic Synthesis | No | Summary of Included Studies | References | |
|---|---|---|---|---|
| | | | Authors/Publication Year/Titles | Relevant Interpretation |
| Integration of Behaviour Change Theories and Mechanisms | 1 | Explicit Theory-Driven: | *Marcus 2022* — Physical activity outcomes from a randomized trial of a theory- and technology-enhanced intervention for Latinas: the Seamos Activas II study | Transtheoretical Model (TTM); Social Cognitive Theory (SCT) |
| | | | *Marcus 2021* — Long-term physical activity outcomes in the Seamos Activas II trial | |
| | | | *Mendoza-Vasconez 2022* — Regular and App-enhanced Maintenance of Physical Activity among Latinas: A Feasibility Study | |
| | | | *Benitez 2016* — Using web-based technology to promote physical activity in Latinas: Results of the Muévete Alabama pilot study | |
| | | | *Marcus 2016* — Using interactive Internet technology to promote physical activity in Latinas: Rationale, design, and baseline findings of Pasos Hacia La Salud | |
| | | | *Marcus 2016* — Pasos Hacia La Salud: a randomized controlled trial of an internet-delivered physical activity intervention for Latinas | |
| | | | *Linke 2019* — Association Between Physical Activity Intervention Website Use and Physical Activity Levels Among Spanish-Speaking Latinas | |
| | | | *Bohlen 2024* — Six-Month Outcomes of a Theory- and Technology-Enhanced Physical Activity Intervention for Latina Women (Pasos Hacia La Salud II): Randomized Controlled Trial | |
| | | | *Ash 2024* — Pasos Hacia La Salud II: A Superiority RCT Utilizing Technology to Promote Physical Activity in Latinas | |
| | 2 | Theory-Aligned but Not Explicitly Stated | *Choi 2021* — A Pilot Study to Promote Active Living among Physically Inactive Korean American Women | No formal theoretical framework was used. Behavioural counselling and education are described, but no specific behaviour change theory (SCT/TTM) is named as the formal framework. |
| | | | *Nguyen 2024* — An App-Based Physical Activity Intervention in Community-Dwelling Chinese-, Tagalog-, and Vietnamese-Speaking Americans | No formal theoretical framework was used. Focuses on translation and feasibility of an app-based intervention; theory is not explicitly framed around SCT or TTM. |
| | | | *Figueroa 2021* — Conversational Physical Activity Coaches for Spanish- and English-Speaking Women: A User Design Study | No formal theory explicitly cited, but design grounded in behavioral reinforcement principles, self-efficacy concepts, and user-centered design (UCD) methodology. Grounded in motivational interviewing principles (open questions, reflective listening, affirmations) to shape coach dialogue; SCT/TTM are not presented as primary frameworks. |
| | | | *Bender 2016* — Designing a Culturally Appropriate Visually Enhanced Low-Text Mobile Health App Promoting Physical Activity for Latinos: A Qualitative Study | No formal theory explicitly cited. |
| | | | *Chee 2016* — Practical Issues in Developing a Culturally Tailored Physical Activity Promotion Program for Chinese and Korean American Midlife Women: A Pilot Study | No formal theory explicitly cited, but the intervention and instruments were influenced by: Theory of Planned Behaviour (TPB), Health Belief Model (HBM). This is evident in the emphasis on attitudes, subjective norms, perceived control, and behaviour intention. |
| | | | *Shin 2022* — A Technology-Enhanced Physical Activity Intervention: A Feasibility Study | No formal theory explicitly cited. |
| Long-time intervention | | | Seamos Activas II focus on deeper integration of behaviour change theory and long-term maintenance testing. Addressing the paucity of technology-supported home-based PA interventions for Latinas and expanding the literature by shifting the focus to achieving the long-term, health-enhancing levels of physical activity needed to resolve related health disparities in this community. | |
| | | Seamos Activas II | *Benitez 2020* — Design and rationale for a randomized trial of a theory- and technology- enhanced physical activity intervention for Latinas: The Seamos Activas II study | The first article presented the design and rationale of the trial, outlining its theoretical foundations in Social Cognitive Theory and the Transtheoretical Model, and describing the six-month randomized controlled design with an additional maintenance phase. |
| | | | *Marcus 2022* — Physical activity outcomes from a randomized trial of a theory- and technology-enhanced intervention for Latinas: the Seamos Activas II study | The second reported the 6-month outcomes, demonstrating that technology-enhanced strategies significantly increased moderate-to-vigorous physical activity (MVPA) and step counts among Latina women during the first six months. |
| | | | *Marcus 2022* — Long-term physical activity outcomes in the Seamos Activas II trial | The third extended this analysis to the maintenance period, showing that women in the intervention arm sustained higher levels of PA than controls at 12 months, although engagement declined over time, highlighting challenges in long-term adherence. |
| | | | *Mendoza-Vasconez 2022* — Regular and App-enhanced Maintenance of Physical Activity among Latinas: A Feasibility Study | This was not part of the original RCT maintenance phase. Instead, it was a separate 3-month follow-on feasibility study, recruiting participants who had already completed Seamos Activas II. |
| | | *Pasos Hacia La Salud I and II:* focused on scalability and long-term digital engagement (e.g., website use, SMS response, sustainability of activity). | | |
| | | Pasos Hacia La Salud I: established feasibility and effectiveness. | *Benitez 2016* — Using web-based technology to promote physical activity in Latinas: Results of the Muévete Alabama pilot study | The goals of the current study were to examine the feasibility and acceptability of using interactive Internet technology to promote physical activity among Latinas in Alabama. Served as a precursor to Pasos I. |
| | | | *Marcus 2015* — Using interactive Internet technology to promote physical activity in Latinas: Rationale, design, and baseline findings of Pasos Hacia La Salud | Presented study rationale, cultural tailoring, and baseline participant data. Established the framework for testing internet-delivered PA interventions. |
| | | | *Marcus 2016* — Pasos Hacia La Salud: a randomized controlled trial of an internet-delivered physical activity intervention for Latinas | Reported efficacy of an internet-delivered PA intervention, showing improvements in MVPA and step counts. Provided rigorous outcome evidence. |
| | | Pasos Hacia La Salud II: refined theory, technology, and sustainability focus. | *Linke 2019* — Association Between Physical Activity Intervention Website Use and Physical Activity Levels Among Spanish-Speaking Latinas | Greater engagement with the website, especially logging activity, setting goals, and reading tips, was significantly associated with higher MVPA; this shows that engagement is a critical mechanism of digital intervention success. |
| | | | *Bohlen 2024* — Six-Month Outcomes of a Theory- and Technology-Enhanced Physical Activity Intervention for Latina Women (Pasos Hacia La Salud II): Randomized Controlled Trial | Both groups improved PA at 6 months; no overall between-group difference. However, among low-active women (some baseline PA), the enhanced intervention was significantly more effective (higher MVPA, more meeting guidelines). No benefit observed among inactive women at 6 months. |
| | | | *Ash 2024* — Pasos Hacia La Salud II: A Superiority RCT Utilizing Technology to Promote Physical Activity in Latinas | At 18 and 24 months, the enhanced arm showed higher MVPA (both self-report and accelerometer) than the original; effect moderated by baseline activity (low-active benefited earlier, inactive women benefited by 24 months). |

E  Appendix E: Summary of Recommendations from Included Studies

| Topics | Recommendation | References | | | | |
|---|---|---|---|---|---|---|
| | | Authors/Publication Year/Titles | Key Outcomes of the Intervention | Research Insights | Future Research Directions | |
| Cultural and Linguistic Adaptation for Accessibility | Prioritizing language accuracy, cultural sensitivity, effective recruitment and retention strategies, and realistic implementation planning to ensure feasibility and sustainability. | Chee 2016 Practical Issues in Developing a Culturally Tailored Physical Activity Promotion Program for Chinese and Korean American Midlife Women: A Pilot Study | 1. Low response, interest, and retention – dropout rate of ~30%; participants found the program burdensome and reimbursement insufficient. 2. Recruitment and retention barriers – inadequate incentives, time burden, and perceived judgment during coaching reduced engagement. | 1. Bilingualism can lead to some degree of deviance in translated meanings. 2. Cultural sensitivity was essential in developing and implementing the culturally tailored Web-based intervention. 3. Another culture-related issue was the women's own perception of their level of physical activity. | Future research should address cultural attitudes: explore and integrate group-specific cultural perspectives (e.g., aging, stigma around depression) into program design. | |
| | Designing culturally relevant, visually accessible, and low-text digital tools to overcome literacy barriers and promote sustained physical activity | Bender 2016 Designing a Culturally Appropriate Visually Enhanced Low-Text Mobile Health App Promoting Physical Activity for Latinos: A Qualitative Study | 1. Both promotores and healthcare providers (HCPs) supported the idea of a visually enhanced low-text (VELT) mHealth app as an alternative to text-heavy materials. | 1. Low health literacy is a barrier, many Latino clients struggle with text-based health information, pointing to the need for visual, simple, and culturally tailored communication. 2. Cultural values like familismo (family cohesion) strongly shape what types of visuals and activities resonate. 3. Relatable and diverse visuals (different ages, body types, family roles) can improve engagement and motivation. 4. Providers face systemic barriers (limited consultation time, lack of appropriate materials) that digital tools could help address. | Future research should develop and test culturally adapted VELT mHealth apps that combine visuals, simple text, and culturally meaningful activities. Such tools have the potential to improve health literacy, support behavior change, and reduce health disparities among populations with low literacy skills. | |
| | Designing culturally and linguistically adapted, theory-driven Internet interventions that reduce barriers, enhance personalization, and clarify PA maintenance for long-term sustainability. | Benitez 2016 Using interactive Internet technology to promote physical activity in Latinas: Rationale, design, and baseline findings of Pasos Hacia La Salud | 1. Cultural and linguistic tailoring: The intervention was adapted to Spanish language and Latina cultural norms, including gender roles and caregiving barriers. 2. Theory-driven design: Grounded in Social Cognitive Theory (SCT) and the Transtheoretical Model (TTM), incorporating stage-matched tailoring, self-monitoring, goal-setting, and feedback. 3. Low baseline PA and psychosocial readiness: Most participants were in contemplation/preparation stages, with very low MVPA, low self-efficacy, and limited social support, highlighting the high-need context. | 1. Tailored interventions such as this have shown efficacy over non-tailored interventions 2. Convenience matters: Web-based access aligned well with Latinas' family and caregiving responsibilities, reducing barriers like childcare and transportation. 3. Role of acculturation and gender norms: Latina women's cultural responsibilities (e.g., marianismo, family caretaking) were central barriers shaping PA adoption. | Future interventions should continue to leverage culturally and linguistically tailored, theory-based digital strategies while addressing key barriers such as low self-efficacy, limited social support, and measurement challenges. Enhancing personalization, integrating more interactive features, and exploring the role of mobile technologies as access improves could strengthen long-term engagement. Importantly, developing standardized approaches to measure and define PA maintenance will be critical for evaluating sustainability across diverse Latina populations. | |
| Personalization and Adaptive Design | Advancing digital health interventions that are personalized and culturally relevant to sustain engagement in underserved populations. | Figueroa 2021 Conversational Physical Activity Coaches for Spanish- and English-Speaking Women: A User Design Study | 1. Low-income Spanish- and English-speaking women are interested in using chatbots to improve their physical activity and general health and feel supported by these tools. 2. Chatbots were perceived as acceptable, but long-term engagement risks were noted. | 1. Digital literacy, app familiarity, and privacy are major concerns that must be addressed for adoption. 2. Co-creation with the target group and collaboration with community partners will improve usability and effectiveness. 3. Social and family inclusion: Many participants wished for a social component in the chatbot; Spanish speakers in particular wanted family members, including children, to engage with the tool. 4. Engagement may wane if chatbot functions and recommendations remain static, leading to boredom. | Future research should design and test culturally tailored, adaptive chatbot interventions that prioritize personalization, safety, and equity, while integrating social and family-based support to sustain engagement in low-income bilingual populations. | |
| | Designing adaptive, theory-based digital platforms that sustain engagement through personalized feedback, dynamic content | Linke 2019 Association Between Physical Activity Intervention Website Use and Physical Activity Levels Among Spanish-Speaking Latinas | 1. Higher engagement linked to better PA outcomes: More log-ins and time spent on the website predicted significantly higher MVPA at 12 months (both objectively and self-reported). | 1. Online community increased intervention adherence, and those who had little support at baseline used and benefited more from this feature compared with those who already had a supportive social network. 2. Engagement declines over time: Website use peaked early and dropped during maintenance, mirroring other digital health interventions. 3. Feature-specific engagement matters: Goal-setting, personalized feedback, and progress reports drove behavior change more than social support/message boards. 4. Usability challenges: Features like MapMyWalk were both valued and difficult to use, revealing a tension between functionality and accessibility. | Future interventions should prioritize sustained engagement strategies by making digital features more adaptive, dynamic, and user-friendly over time. Design should emphasize high-impact features (goal-setting, personalized feedback, progress tracking) while simplifying navigation and reducing digital burden. | |
| | Expanding culturally tailored mobile apps with social features and healthcare linkages to sustain PA engagement | Nguyen 2024 An App-Based Physical Activity Intervention in Community-Dwelling Chinese-, Tagalog-, and Vietnamese-Speaking Americans | 1. Demonstrated high feasibility and acceptability of a culturally tailored, multilingual app-based PA program. 2. Participants reported positive experiences, including an increased motivation to walk and the enjoyment of being able to monitor their physical activity. | 1. This study builds on prior evidence from a culturally adapted, multicomponent intervention for Filipino Americans with type 2 diabetes, which demonstrated feasibility and potential efficacy. 2. By extending this work to Chinese-, Tagalog-, and Vietnamese-speaking communities, it highlights the need for multilingual tailoring to engage diverse Asian American subgroups. | Future work should integrate customizable options, add social features (eg, social comparisons and challenges), and incorporate communication with health care clinicians, which may be beneficial to increase adherence to the intervention components. | |
| | Advancing stage-sensitive, theory-driven digital interventions with tailored intensity to support both initiation and maintenance of PA. | Bohlen 2024 Six-Month Outcomes of a Theory- and Technology-Enhanced Physical Activity Intervention for Latina Women (Pasos Hacia La Salud II): Randomized Controlled Trial | 1. For low-active Latina women, the enhanced intervention was more effective at increasing PA. 2. Inactive women saw no extra benefit: For women who were completely inactive at baseline, enhanced features did not provide additional gains over the original program. 3. Additional tailored intervention enhancements may be necessary to increase PA for inactive Latina women. | 1. Baseline activity moderated effects: Intervention effectiveness varied by stage of readiness—low-active participants benefited most. 2. Enhancements supported engagement: Added text messages, dynamic features, and extra phone calls strengthened SCT constructs (self-efficacy, enjoyment, social support). 3. Challenge with inactive participants: Additional strategies are required to help completely inactive women initiate activity. | Future work should focus on stage-sensitive intervention designs, providing more intensive motivational and behavioral support for completely inactive women, while leveraging lighter enhancements for low-active participants. Exploring mobile apps, push notifications, and wearables may improve engagement and initiation. To ensure comparability across studies, researchers should also work toward defining and standardizing physical activity maintenance as a long-term outcome. | |
| Structured and Equitable Social Support | Leveraging culturally tailored and socially supportive digital platforms to promote scalable and sustainable physical activity engagement | Choi 2021 A Pilot Study to Promote Active Living among Physically Inactive Korean American Women | 1. Increased self-efficacy for exercise knowledge 2. Improved perceived social encouragement and use of the physical environment 3. Reduced perceived barriers to exercise 4. Reduced mental distress | 1. Women who were more active in the online social community already walked more at the start, and this higher level of activity persisted. 2. Women who engaged less in the community consistently walked less, suggesting they may need extra encouragement or support. 3. Having relatable role models may be especially important for physically inactive women to start and maintain physical activity. | Future work should test whether online social communities improve both adoption and long-term maintenance of physical activity. Studies ought to run longer with clear follow-ups, and replace resource-heavy one on one onboarding with a group session plus brief remote check-ins and tailored text messages to boost scalability. | |
| | Designing multi-component, culturally tailored interventions that combine accessible digital tools with personalized human support is critical to reducing disparities and sustaining physical activity among underserved populations. | Collins 2019 Efficacy of a multi-component intervention to promote physical activity among Latino adults: A randomized controlled trial | 1. For both groups, there was a statistically significant within-group increase in aerobic and strength training exercise behaviors. 2. There were no significant differences between groups at three months in quality of life or exercise behavior scores. | 1. Multi-component design works: Combining text messaging (stage-tailored encouragement) with motivational interviewing (MI) phone calls created a stronger boost in steps compared to handouts alone. 2. Human contact is valued: Participants rated MI phone calls slightly higher than text messages, showing the value of brief but personal interaction. | Future studies should test this multicomponent model in larger and longer trials to determine sustained effects. Expanding the intensity or personalization of motivational interviewing, as well as integrating text messages culturally and linguistically tailored, may further enhance engagement. Designs should also address health literacy and education differences to ensure equitable impact. | |
| | Developing culturally and linguistically adapted, theory-driven digital interventions that embed structured and equitable social support mechanisms to sustain physical activity. | Benitez 2016 Using web-based technology to promote physical activity in Latinas: Results of the Muévete Alabama pilot study | 1. Increase in PA. Self-reported MVPA increased significantly from 12.5 min/week at baseline to 67.5 min/week after one month. 2. Psychosocial improvements. Significant gains in self-efficacy, cognitive processes of change, and behavioral processes of change. 3. Stages of change progression. Nearly half (45.8%) of participants advanced at least one stage of readiness for PA. 4. High feasibility and satisfaction. Retention was strong (88%), and all participants who completed feedback reported gaining knowledge, motivation, and satisfaction with the program. | 1. the culturally and linguistically adapted, theory-(Social Cognitive Theory and the Transtheoretical Model) Internet-based physical activity intervention for Latina adults produced significant increases in self-reported moderate-to-vigorous intensity PA, processes of change, and self-efficacy for physical activity. 2. The intervention allowed participants to select their own support partner (friend/family) to share the website and pedometer. • This worked only if women already had supportive social networks. • For those with weak or absent networks, this strategy created a barrier, which may explain the lack of significant increases in perceived social support. | Future interventions should strengthen cultural tailoring, family integration, and digital usability to further support PA adoption among Latinas. Importantly, future research should also explore standardized definitions of PA maintenance and integrate objective and self-report measures to better evaluate sustainability over time. In addition, future interventions should move beyond relying solely on participants' existing social networks and instead provide structured or facilitated opportunities to connect with peers or assigned partners. Embedding social matching, group-based activities, or community health worker facilitation into digital interventions could ensure that even women with limited social networks receive meaningful social support. | |
| Long-Term Maintenance and Habit Reinforcement | Reinforcing long-term PA maintenance through theory-driven, technology-enhanced strategies that sustain newly formed habits with minimal but continuous support. | Marcus 2021 Long-term physical activity outcomes in the Seamos Activas II trial | 1. Gains in PA at six months were maintained at 12 months for both interventions. 2. Enhancements (theory, texting) can help high risk populations sustain PA. 3. Participants in the Enhanced arm were more likely to meet PA guidelines (≥150 min/week of MVPA) at 12 months, suggesting low-touch and low-burden theoretical and technological enhancements may help Latinas to continue meeting PA guidelines during the maintenance phase of PA interventions. | 1. Evidence-based behavior change tools work: Even minimal "maintenance phase" contact (low-touch texts, brief phone call) helped participants sustain PA. 2. Theoretical and technological enhancements matter: Targeting SCT constructs (self-efficacy, enjoyment, social support, outcome expectancies) supported better adherence to guidelines. 3. Cultural subgroup differences: Mexican-American participants in San Diego showed stronger outcomes than Dominican/Cuban groups in the prior Seamos Saludables study, suggesting cultural context shapes efficacy. 4. Long-term PA maintenance is rare but achievable: Few Latina-focused trials measure 12-month outcomes, making this evidence particularly valuable. | Future interventions should continue to use evidence-based behavior change tools during the maintenance phase to reinforce habits, even with minimal contact. Integrating theoretical frameworks and technological enhancements can further assist in sustaining PA and consistently meeting guidelines over the long term. | |
| | Designing culturally relevant, user-friendly digital supports that sustain physical activity habits while advancing consensus on what constitutes PA maintenance. | Mendoza-Vasconez 2022 Regular and App-enhanced Maintenance of Physical Activity among Latinas: A Feasibility Study | 1. Sustained PA gains overall: MVPA dropped after the 12-month intervention but remained much higher than pre-intervention levels (163 min/week vs. 17 min/week at baseline). 2. Enhanced maintenance group: Reported a smaller decline in MVPA compared to the regular group (−29 vs. −85 minutes/week), though not statistically significant. | 1. Our results highlight the need for a consensus and an individual-centered definition of PA maintenance. 2. App feasibility and acceptability: Commercial apps (Fitbit, Nike+) were moderately used and valued for self-monitoring, visual appeal, and overcoming barriers (e.g., weather). 3. Habit formation reduces reliance on apps: Some women maintained PA without apps, suggesting that once habits are internalized, digital prompts become less essential. 4. Barriers: Lack of tech skills, frustration with log-ins, and apps not offering culturally relevant activities (like Zumba) reduced engagement. | Future interventions should move beyond simply adding apps and instead design flexible, culturally resonant, and user-friendly digital supports that enhance long-term PA maintenance. This includes offering a variety of app options that align with participants' cultural preferences (e.g., activities like Zumba or energetic dance), simplifying usability to overcome digital literacy barriers, and integrating apps earlier in interventions to support habit formation. Importantly, to advance the field, it is necessary to reach a consensus on what constitutes physical activity maintenance, so that future interventions can be designed, evaluated, and compared more effectively. | |
| | Integrating lifestyle components, reinforcing self-efficacy, and embedding family/peer support are essential for sustaining physical activity in CALD women | Shin 2022 A Technology-Enhanced Physical Activity Intervention: A Feasibility Study | 1. High feasibility and acceptability: 100% retention, 93% adherence, and 88.4% satisfaction demonstrated strong engagement and practicality. 2. Behavioral improvements: Significant increases in self-efficacy, moderate PA, and daily step counts. 3. Health outcomes: Significant reductions in BMI and systolic blood pressure; reductions in prediabetes rates. 4. Many participants only discovered prediabetes or hypertension risks during the study | 1. Self-efficacy is central. Gains in confidence strongly correlated with increased PA, reinforcing its importance as a behavioral driver. 2. Technology plus personal contact matters. Combining Fitbit self-monitoring with text reminders and small group support created a motivational synergy. 3. Social support is complex. Despite intervention efforts, family and friend support did not significantly increase—possibly due to competing family demands in midlife. 4. Health risks were under-recognized. About 21% of participants learned they had prediabetes or hypertension for the first time, highlighting limited healthcare access and awareness in this population. | Future interventions should combine physical activity with other lifestyle factors such as diet, while also strengthening social support through family or peer involvement. Reinforcing self-efficacy with achievable goals and providing longer-term follow-up will be key to sustaining engagement and improving health outcomes. | |

| Topics | Recommendation | References | | | |
|---|---|---|---|---|---|
| | | Authors/Publication Year/Titles | Key Outcomes of the Intervention | Research Insights | Future Research Directions |
| | Developing culturally tailored strategies to reduce attrition and sustain long-term engagement in mHealth interventions | Rowland 2022<br>*Feasibility, Usability and Acceptability of a mHealth Intervention to Reduce Cardiovascular Risk in Rural Hispanic Adults* | 1. Among those who received mHealth support in the intervention group, the usability of the MyFitnessPal app and smart scale to self-monitor daily steps, weight, and calorie intake was high and increased over time. | 1. The intervention group received in-person visits with a bilingual nurse practitioner and a technology support person, which may have facilitated the development of self-monitoring skills (weight, calorie intake, and physical activity).<br>2. mHealth can be leveraged to promote public health and help patients self-manage cardiovascular risk factors, particularly in the absence or limited supply of health care providers in rural communities. | Future research should evaluate and support participants' baseline technology skill level, provide training if needed, and use a phone call or SMS text message follow-ups as a strategy to minimize attrition. In addition, future studies should develop and test culturally tailored strategies to reduce attrition and sustain long-term engagement with mHealth interventions among Hispanic/Latino adults. |
| Theory-Driven and Multi-Channel Enhancement | Enhancing culturally tailored interventions with interactive digital supports to build accountability, enjoyment, and social support for sustaining physical activity. | Benitez 2020<br>*Design and rationale for a randomized trial of a theory- and technology-enhanced physical activity intervention for Latinas: The Seamos Activas II study* | 1. Participants were highly inactive: At baseline, objectively measured MVPA was very low (≈34–46 min/week), with self-reported MVPA near zero for most.<br>2. Low psychosocial readiness: Most women were in precontemplation or contemplation stages of change, with generally low self-efficacy and reliance on cognitive (not behavioral) processes of change. | 1. Theory refinement: More rigorous targeting of Social Cognitive Theory constructs (e.g., enjoyment, outcome expectations, social support) may be critical to promoting PA.<br>2. Interactivity matters: Participant feedback from prior studies highlighted the need for greater accountability and immediacy, leading to the integration of interactive texting.<br>3. Family and friend support as mediators: Evidence from earlier trials suggested that family/friend support mediated sustained PA; the new design built in more opportunities to leverage this support. | Future studies should address interactivity and accountability through text messaging, and more rigorously targeting theoretical constructs may be key to helping Latinas achieve nationally recommended PA levels and thereby reducing health disparities. |
| | Enhancing theory-driven, multi-channel digital interventions with tailored intensity to sustain physical activity and clarify definitions of PA maintenance | Marcus 2016<br>*Pasos Hacia La Salud: a randomized controlled trial of an internet-delivered physical activity intervention for Latinas* | 1. Internet-delivered, individually tailored intervention achieved even larger increases in physical activity than the print-based version<br>2. Baseline PA moderated effects: Low-active participants benefitted earlier (by 12–18 months), while completely inactive participants only showed benefits by 24 months.<br>3. High retention: 81% of participants completed the 24-month follow-up, showing the feasibility of long-term technology-based interventions. | 1. Enhancements worked: Adding text messaging, gamification features, persuasive elements (rewards, reminders), and extra support extended PA gains beyond one year.<br>2. Stage of readiness matters: Inactive women needed longer to benefit, reflecting Transtheoretical Model stages (precontemplation vs. preparation/action).<br>3. Technology integration helps: Combining website + text messaging increased engagement and sustained usage, compared with website-only approaches.<br>4. Theory-driven foundation: SCT and TTM guided tailoring, self-monitoring, and stage-matched feedback, reinforcing the value of theory-based designs. | Future interventions should refine multi-channel, theory-driven digital strategies that combine web, text, and gamified features to sustain long-term PA. Design should account for differences in baseline readiness by providing more intensive early-stage support for inactive participants and lighter-touch reinforcement for low-active women. Importantly, advancing the field will require consensus on what constitutes PA maintenance, so outcomes can be consistently defined and compared across studies. |
| | Scaling culturally tailored, theory-driven, technology-enhanced interventions to sustain physical activity and reduce health disparities. | Marcus 2022<br>*Physical activity outcomes from a randomized trial of a theory- and technology-enhanced intervention for Latinas: the Seamos Activas II study* | 1. Increased PA across both arms: Both the Original and Enhanced interventions significantly increased MVPA at 6 months (accelerometer and self-report).<br>2. Higher odds of meeting guidelines: 57% of Enhanced group vs. 44% of Original group met national aerobic PA guidelines at 6 months | 1. Theory + technology boosts outcomes: Adding SCT-focused tailoring and text messaging improved accountability and enjoyment, helping more Latinas achieve PA guidelines.<br>2. Home-based, low-burden interventions are effective: Delivering interventions by mail + texts reduced barriers like transportation and caregiving demands. | Future studies should test the scalability of home-based, culturally tailored, and technology-enhanced PA programs in diverse Latina subgroups. Longer-term follow-ups are needed to evaluate sustained adherence and cardiometabolic outcomes. Research should also explore tailoring interventions to different cultural contexts and family norms, while refining digital tools to address both behavioral support and equity in access. |
| | Refining stage-sensitive, theory-driven digital interventions with adaptive personalization. | Ash 2024<br>*Pasos Hacia La Salud II: A Superiority RCT Utilizing Technology to Promote Physical Activity in Latinas* | 1. Both groups improved MVPA: Both the original and enhanced interventions led to sustained increases in MVPA over 24 months.<br>2. Baseline PA moderated outcomes: Low-active participants benefitted earlier (12–18 months), while completely inactive women showed significant gains only by 24 months.<br>3. High retention: 81% of participants completed the 24-month follow-up, demonstrating the feasibility of long-term digital delivery. | 1. Enhancements worked: Adding text messaging, gamification (tokens, medals), persuasive elements (reminders, rewards), and staff-facilitated social support improved engagement and outcomes.<br>2. Stage of readiness matters: Inactive participants required more time and intensity to see benefits, while low-active participants responded sooner.<br>3. Theory-driven + technology integration: SCT and TTM underpinned tailoring, stage-matched feedback, and motivational strategies, confirming the value of theory-based design.<br>4. Persuasive technology potential: Combining website + text messaging created multiple "touch points" that sustained adherence better than static content alone. | Future work should refine stage-sensitive, theory-driven, multi-channel interventions that deliver differentiated support for inactive and low-active women. Incorporating adaptive personalization, sustained engagement strategies (gamification, reminders, rewards), and broader technology integration can help reinforce long-term PA maintenance. Importantly, future research should also work toward establishing consensus on how PA maintenance is defined and measured to allow consistency across trials. |